# A Theory of Extrasolar Giant Planets

D. Saumon[1,2], W.B. Hubbard[2], A. Burrows[3], T. Guillot[2], J.I. Lunine[2], and G. Chabrier[4]

## ABSTRACT

We present a broad suite of models of extrasolar giant planets (EGP's), ranging in mass from 0.3 to 15 Jupiter masses. The models predict luminosity (both reflected and emitted) as a function of age, mass, deuterium abundance and distance from parent stars of various spectral type. We also explore the effects of helium mass fraction, rotation rate and the presence of a rock-ice core. The models incorporate the most accurate available equation of state for the interior, including a new theory for the enhancement of deuterium fusion by electron screening which is potentially important in these low mass objects. The results of our calculations reveal the enormous sensitivity of EGP's to the presence of the parent star, particularly for G and earlier spectral types. They also show a strong sensitivity of the flux contrast in the mid-infrared between parent star and EGP to the mass and age of the EGP's. We interpret our results in terms of search strategies for ground- and space-based observatories in place or anticipated in the near future.

*Subject headings:*

[1] Hubble Postdoctoral Fellow.
[2] Department of Planetary Sciences, University of Arizona, Tucson, AZ 85721 (dsaumon, guillot, hubbard, jlunine@lpl.arizona.edu).
[3] Departments of Physics and Astronomy, University of Arizona, Tucson, AZ 85721 (burrows@cobalt.physics.arizona.edu).
[4] Centre de Recherche Astrophysique de Lyon (UMR 142 CNRS), Ecole Normale Supérieure, 46, Allée d'Italie, 69364 Lyon Cedex 07, France. (chabrier@physique.ens-lyon.fr)

# 1. Introduction

The questions of the existence and properties of other planetary systems represent one of the philosophical centerpieces of modern astrophysics. Planetary systems are an end state of the process of star formation, and the existence of planets imposes an important set of physical constraints on that process. Characteristics of planets provide information about the angular momentum content and evolution of the system, the lifetime of the disks, and the nature of the energetic processes associated with the pre-main sequence star itself.

Giant planets are of particular interest, both because they are most detectable with current and envisioned technologies, and because they pose a puzzle for planet formation. Our prototypes of giant planets are Jupiter and Saturn, which are primarily composed of hydrogen and helium in roughly solar proportions, but which differ in detail from solar composition in that they are enhanced in metals by about one order of magnitude, with a substantial quantity of these heavier elements concentrated toward their centers.

The purpose of this paper is to provide accurate and extensive models of extrasolar giant planet brightnesses and dimensions as a function of age, composition, and mass, both as a guide to what stars (spectral type, age) around which to search for giant planets, and as a tool for interpreting the results of any positive detections. It has long been recognized that essentially the same physics governs the structure and evolution of the suite of electron-degenerate and hydrogen-rich objects ranging from brown dwarfs (at the high mass end) to Jupiters and Saturns (at the low mass end). Except for an initial study by our group (Burrows et al. 1995), no one has quantitatively mapped out the properties of objects between the mass of giant planets in our solar system and the traditional brown dwarfs ($> 10 - 20\ M_J$, where $M_J$ is the mass of Jupiter), which we term extrasolar giant planets (EGP's).

Earlier work generally consists of evolutionary models of planets of $1\ M_J$ and below beginning with Graboske et al. (1975; hereafter GPGO; but see Hubbard, 1977). This work calculates the evolution of the low-mass objects Jupiter and Saturn from an age of $10^7$ years to the present (4.5 Gyr). Working down from higher masses, Grossman and Graboske (1973; GG73) extended their calculations of brown dwarf evolution to as low as $12\ M_J$, but had to limit their study to ages less than about 0.1 Gyr. Black (1980) used the results of GG73 and GPGO to infer simple power-law relations for the variation of luminosity $L$ and radius $R$ as a function of mass $M$ and time $t$. Black's relations are roughly valid for objects close in mass to 1 $M_J$ and close in age to 4.5 Gyr. However, as we discuss below, Black's formulas become very inaccurate at earlier ages and at larger masses.

The scope of this paper is as follows. Our lower mass limit is the mass of Saturn (0.3 $M_J$), and our upper mass limit is 15 $M_J$, which takes us to objects which would generally be considered brown dwarfs. Our baseline models for EGP's are composed of hydrogen and helium with a helium mass fraction $Y = 0.25$ and with a metals mass fraction $Z \approx 0.02$, with the latter playing no significant role in the interior structure. However, we also examine the effect of enhancing metals well above solar composition. Our theory does not include objects similar to the ice giant planets Uranus and Neptune. These belong to a different class of object because they contain minor hydrogen-helium fractions, and their masses and luminosities are an order of magnitude smaller than those of Jupiter and Saturn.

Our theory starts with the assumption that EGP's have been somehow formed from an initially gaseous, high-entropy state. Current theoretical models for forming giant planets require relatively rapid accretion of large amounts of protoplanetary disk gas onto a core of rocky and icy material (Podolak, Hubbard & Pollack 1993). Models of our own protoplanetary disk, or solar nebula, suggest that the timescale for accumulating a giant planet's solid core is long enough that the gaseous accretion stage may potentially be truncated, as appears to have happened for Uranus and Neptune. Although this problem appears to have been solved for Jupiter and Saturn (Lissauer 1995), a potential complication is that the tidal effects of such a growing object may cause a gap in the gas disk to form which truncates or greatly slows accretion (Lin & Papaloizou 1993). Very recent observations (Zuckerman, Forveille & Kastner 1995) suggest that in many protoplanetary systems the bulk of the gas may be dissipated within a few million years and resurrect the timescale problem, since timescales had previously been assumed to be at least several times longer. Further, recent models of protoplanetary disks seem to suggest that Jovian planet formation does not occur within roughly 5 A.U., being relatively insensitive to the spectral type of the cen-tral star for $M \leq 1\,M_\odot$ (Boss 1995). If this is the case, the timescale problem would be common to giant planets forming around protostars of a wide range of masses.

Because of the seeming delicacy in arranging for the successful formation of a giant planet, Wetherill (1993) argued that Jupiters and Saturns may be rarities in planetary systems. If so, this would have interesting consequences for the habitability of terrestrial planets in other systems, since the giant planets have been very effective in sweeping cometary debris from our own solar system. Observations of the $\beta$ Pictoris system suggest orders of magnitude more dust than in our own solar system, even relatively close to the central star (Backman, Gillett & Witteborn 1992), and it is tempting to speculate that this system contains no giant planets to sweep the inner regions clear of small debris.

Walker et al. (1995) have monitored the radial velocity of 21 nearby stars, and find no Jupiter-mass planets on circular orbits of less than 12-year periods. Since this survey is not definitive, it is fair to say that there is controversy in both theory and data about the mode of formation of giant planets and consequently about how common they may be. We believe that this question must be settled observationally, and thus our motive is to provide extensive and quantitative predictions useful for observers who are trying to detect EGP's. The technologies now seem to be sensitive enough for direct detection by imaging from the ground and space (Angel 1994; Burrows et al. 1995), as well as by indirect techniques such as radial velocity (McMillan et al. 1994), precision astrometry (Gatewood 1987), microlensing (Gould & Loeb 1992), and photometric detection of transits of a giant planet (Borucki & Genet 1992).

In the present work, we use updated high pressure thermodynamics, derived from two decades of laboratory and theoretical work (see, e.g., Van Horn & Ichimaru 1993, and Chabrier & Schatzman 1994) and an improved surface boundary condition to construct models ranging from Saturn and Jupiter up through 15 Jupiter masses. In §2, we describe the physics used to model the atmospheres and interiors of these objects. Section 3 presents results of the model, comparing with data on Jupiter and Saturn, and explore the nature of the deuterium burning phase, the effects of varying the flux from the central star, the helium abundance, and the mass of heavy element core. In §4, we utilize our ensemble of new models to predict what extrasolar giant planets should look like, and how and where to target the searches.

## 2. Input physics to the models

We follow the evolution of giant gaseous planets with masses from 0.3 to 15 $M_J$ for 5 Gyr. The models are non-rotating and in hydrostatic equilibrium. We neglect the presence of metals in the interior of the planet but the atmospheric surface boundary condition assumes a solar abundance of heavy elements. The effect of heavy elements concentrated in a central rock-ice core is discussed in §3.3. The calculation is similar to the brown dwarf sequences of Burrows et al. (1993) and to the EGP results of Burrows et al. (1995), where additional information can be found. We upgrade the input physics of Burrows et al. (1993) by using a more accurate equation of state for H/He mixtures, extending the surface boundary condition to lower effective temperatures, and applying state-of-the-art screening corrections to the rate of the $^2\mathrm{D}(p,\gamma)^3\mathrm{He}$ nuclear reaction.

### 2.1. The surface boundary condition: Treatment of the atmosphere

Calculation of the evolution of fully-adiabatic models of EGP's requires a surface condition which can be expressed in the form

$$T_{10} = f(g, T_{\mathrm{eff}}), \qquad (1)$$

where $T_{10}$ is the temperature corresponding to the internal adiabat at a chosen pressure of 10 bars, and $f$ is a function of the surface gravity $g$ and the effective temperature $T_{\mathrm{eff}}$ which is determined from a grid of model atmospheres. Here, $T_{\mathrm{eff}}$ is the effective temperature of a blackbody with the EGP's radius whose thermal luminosity corresponds to the sum of the intrinsic luminosity of the EGP and the absorbed stellar luminosity. The absorbed luminosity is calculated using a Bond albedo of $A = 0.35$ which is characteristic of the giant planets of the solar system.

We have previously determined $f$ for $T_{\mathrm{eff}} \geq 600\,\mathrm{K}$ for the X-model sequence of Burrows et al. (1993). For giant planets at lower values of $T_{\mathrm{eff}}$ and $g$, GPGO determined $f$ in tabular form by integrating model atmospheres in the range

$$20 \leq T_{\mathrm{eff}} \leq 1900\,\mathrm{K} \qquad (2)$$

and for two values of $g$: 40.39 and 2585 $\mathrm{cm\,s^{-2}}$. GPGO took into account Collision-Induced Absorp-

tion (CIA) opacity for $H_2$-$H_2$ and $H_2$-He calculated by Linsky (1969), water opacity from Ferriso et al. (1966) up to 11000 cm$^{-1}$, ammonia opacity (at very low frequencies), and methane opacity (at very low frequencies).

Hubbard (1977) fitted an analytic form for $f$ to the tabulated data of GPGO, with the result

$$T_{10} = 3.36 \, g^{-1/6} T_{\text{eff}}^{1.243} \qquad (3)$$

(all quantities in c.g.s. units). However, this form is accurate only for $T_{\text{eff}} \leq 200$ K. For $T_{\text{eff}} > 200$ K, we find that a better fit is given by

$$T_{10} = 15.86 \, g^{-1/6} T_{\text{eff}}^{0.95}. \qquad (4)$$

Since the GPGO calculations do not extend to gravities greater than the present surface gravity of Jupiter (2600 cm s$^{-2}$), some extrapolation of relations (3) and (4) is required for calculation of the evolution of objects more massive than Jupiter.

As long as CIA is the dominant source of thermal opacity, $f$ is expected to have a weak dependence on $g$. To isolate the $g$-dependence, we can write the equation of hydrostatic equilibrium for the radiative portion of the planetary atmosphere as follows:

$$\frac{dP}{d\tau} = \frac{g}{\kappa}, \qquad (5)$$

where $\tau$ is the optical depth and $\kappa$ is the opacity per unit mass. For CIA, in which the opacity is proportional to the number density of molecules, we can write

$$\kappa = \alpha(T_{\text{eff}}) \frac{P}{T}, \qquad (6)$$

where $\alpha$ is some function of $T_{\text{eff}}$ which does not depend on $P$ or $T$. In the radiative upper atmosphere we write the usual approximation for the $T(\tau)$ relation:

$$T = 2^{-1/4} T_{\text{eff}} \left(1 + \frac{3}{2}\tau\right)^{1/4}. \qquad (7)$$

We assume that the atmosphere becomes convective and thus adiabatic for $\tau \sim 1$. Substituting Eqs. (6) and (7) in Eq. (5) and integrating from $\tau = 0$ to $\tau = 1$ yields

$$P_{\tau=1}^2 \propto gT_{\text{eff}}/\alpha(T_{\text{eff}}). \qquad (8)$$

Hubbard (1977) assumed that in the adiabatic portion of the atmosphere, which commences at $\tau > 1$, $P \propto T^3$, or $P_{\tau=1}^2 = P^2(T_{\tau=1}^6/T^6)$. Thus, at a fixed pressure of 10 bars, Eq. (8) leads to $T_{10} \propto g^{-1/6}$, in agreement with Eqs. (3) and (4).

For hydrogen-helium adiabats in the temperature range considered ($100 \leq T \leq 1000$ K), the adiabatic relation is more accurately written as $P \propto T^{3.3}$, which leads to a slightly weaker dependence of $T_{10}$ on $g$: viz. $T_{10} \propto g^{-1/6.6}$; we ignore this complication considering the crudity of the other approximations.

We adopt Eqs. (3) and (4) for $T_{\text{eff}} \leq 300$ K, and we use the X grid of surface conditions (Burrows et al. 1993) for $T_{\text{eff}} \geq 600$ K. Simple interpolation is used to determine boundary conditions for objects which lie between the two ranges. Figure 1 shows, for six different values of $g$, the X surface conditions, and the surface conditions (3–4), along with the interpolation region. As the figure indicates, the actual $g$-dependence of $T_{10}$ may be somewhat steeper than $g^{-1/6}$ for $T_{\text{eff}} \leq 600$ K and for surface gravities greater than $10^4$ cm s$^{-2}$. For improved results it will be eventually necessary to calculate atmosphere models in this range.

### 2.2. Thermodynamics of the interior

Simple physical arguments and detailed calculations indicate that stars with masses below $\sim 0.3 \, M_\odot$ (or $\sim 300 \, M_J$) have fully convective interiors and that this state persists through the regime of brown dwarfs down to giant planets like Saturn. It follows from the high convective efficiency found in these low-$T_{\text{eff}}$ objects that the interior structure is adiabatic. Models are obtained by integrating the equation of hydrostatic equilibrium along adiabats generated with the equation of state (EOS) of Saumon, Chabrier & Van Horn (1995, hereafter SCVH) which was developed for applications to very-low mass stars, brown dwarfs and giant planets. In these relatively dense and cool objects, nonideal effects dominate the physics of the EOS, particularly at densities above $\approx 0.1 \, \text{g cm}^{-3}$ where neutral particles (e.g. $H_2$) strongly repel each other and ultimately become pressure-ionized to form a strongly-coupled plasma. These effects are carefully accounted for in the SCVH EOS, which is the most accurate available for these objects.

The SCVH EOS reproduces all relevant experimental results very well except for the new measurements by Holmes, Ross & Nellis (1995) on shock compressed deuterium which disagree with the SCVH EOS in the regime of pressure dissociation of $H_2$ molecules. The new data suggest a larger degree of dissociation than predicted by SCVH. This effect has potentially significant consequences for the interior of giant planets (Nellis, Ross & Holmes 1995) but it is not included in the present work. Modifications of the EOS to bring it into agreement with the new measurements are currently under way.

Finally, we adopt a helium mass fraction of $Y = 0.25$ for the interior models. The sensitivity of the models to the helium mass fraction is discussed in §3.3. We find that the calculated emissions of EGP's (§4.2) are barely affected when using a value of $Y = 0.28$.

### 2.3. Screening Correction to the $^2\mathrm{D}(p,\gamma)^3\mathrm{He}$ Reaction Rate

Marginal ignition of deuterium via the reaction $^2\mathrm{D}(p,\gamma)^3\mathrm{He}$ occurs in objects in the EGP mass range. The first study of this topic was by GG73, who found a deuterium main sequence starting at 0.012 $M_\odot$ (12 $M_J$). A deuterium main sequence is obtained when the luminosity of the object is entirely provided by the burning of deuterium.

We have updated and expanded upon GG73 in several ways. First, the initial deuterium abundance for our objects, which is taken to be the protosolar value, is D/H= $2 \times 10^{-5}$, while GG73 chose D/H= $1.9 \times 10^{-4}$, which is the (enriched) terrestrial ratio of deuterium to hydrogen. Second, GG73 did not carry their calculations below $T_{\mathrm{eff}} = 1260$ K, while our interpolation relation (§2.1) permits us an essentially unlimited range of $T_{\mathrm{eff}}$. Third, our equation of state include extensive treatment of nonideal behavior, and is quantitatively applicable to all masses in the range 0.3 to 15 $M_J$, and higher. Finally, GG73 calculated thermonuclear reaction rates for $^2\mathrm{D}(p,\gamma)^3\mathrm{He}$ taking ion screening into account, but not electron screening. In such low-mass objects, partially degenerate electrons in metallic hydrogen effectively shield the protons and deuterons, making it easier for them to overcome their mutual Coulomb barrier.

The global enhancement factor $\exp[H(0)]$ thus includes ionic and electronic contributions. The quantity $H(0)$ for reactions between the two charges $Z_1$ and $Z_2$ is exactly equal to the difference between the excess (non-ideal) free energies $F^{\mathrm{ex}}$ in the plasma before and after the reaction (Jancovici 1977):

$$H(0) = \; F^{\mathrm{ex}}(\Gamma_1, Z_1; \Gamma_e, r_s) + F^{\mathrm{ex}}(\Gamma_2, Z_2; \Gamma_e, r_s) \\ - F^{\mathrm{ex}}(\Gamma_{12}, Z_{12}; \Gamma_e, r_s) \quad (9)$$

where $\Gamma_i = \Gamma_e Z_i^{5/3}$ is the ionic coupling parameter for the ion of charge $Z_i$, and $Z_{ij} = Z_i + Z_j$ is the charge of the fused pair. The electronic coupling parameter is

$$\Gamma_e = e^2/a_e k_B T, \quad (10)$$

and the coupling parameter of the quantum electrons is

$$r_s = a_e/a_0 \quad (11)$$

where $e$ is the charge of the electron, $a_e$ is the mean inter-electron spacing ($4\pi a_e^3/3 = n_e^{-1}$), $k_B$ is the Boltzmann constant, and $a_0$ is the Bohr radius. The ionic contribution is obtained directly from Eq. (9) with the most recently determined fits for the OCP free energy (DeWitt, Chabrier & Slattery 1995). The electronic contribution has been calculated using a polarization potential, i.e. the difference between the bare Coulomb potential and the screened-Coulomb potential, which takes into account the afore-mentioned electron polarization in the interionic potential through the electron dielectric function (Chabrier 1990):

$$V_{ij}^{\mathrm{pol}}(r) = \frac{Z_i Z_j e^2}{2\pi^2} \int \frac{1}{k^2} \left[ \frac{1}{\epsilon(k, r_s, T)} - 1 \right] \exp(i\vec{k} \cdot \vec{r}) \, d\vec{k} \quad (12)$$

Here $\epsilon(k, r_s, T)$ is the electron dielectric function as a function of spatial wavenumber $k$, electron coupling parameter, $r_s$, and (finite) temperature $T$, which takes into account the electron-electron correlations beyond the RPA approximation through the so-called local field correction.

Equation (9) relies on the so-called linear-mixing rule, where the free energy of the mixture is given by the linear interpolation of the free energies of the pure components. The accuracy of this approximation was demonstrated initially for the bare Coulomb potential (Brami, Hansen & Joly 1979) and has been verified for the screened Coulomb potential (Chabrier & Ashcroft 1990).

A complete presentation of the present formalism, and its application to the nuclear reactions of light elements in low-mass stars will be given in a forthcoming paper (Chabrier 1995).

### 3. Structure and Evolution

#### 3.1. The Deuterium-burning Phase

Several models straddling the limiting mass for deuterium burning are presented in Fig. 2. Panel (a)

shows the luminosity as a function of time, for models with masses from $10\,M_J$ to $15\,M_J$ in steps of $1\,M_J$ (masses increase upward). The solid dot is the lowest-mass ($12\,M_J$) model of GG73, during the phase on the deuterium-burning main sequence which GG73 find for an elevated D/H value. Panel (b) shows $f_N$, the fraction of the luminosity derived from $^2D(p,\gamma)^3\mathrm{He}$. The deuterium main sequence is defined by $f_N = 1$. In Fig. 2, the solid curves are calculated with the full thermonuclear screening corrections, while the dotted curves are calculated with the older ion-only screening corrections. The electron screening theory takes into account electrons at finite temperature, but the effect of thermal corrections to the electron distribution is very slight and a $T = 0$ theory for electron screening appears to be adequate. The heavier curves in Fig. 2 show the transitional model of $13\,M_J$, in which D/H declines from an initial value of $2 \times 10^{-5}$ to a final value of $1.5 \times 10^{-5}$. For comparison, with ion-only screening, the final deuterium abundance in the $13\,M_J$ model is $1.7 \times 10^{-5}$.

Panel (c) shows curves of D/H vs. mass for various times, with the final values of D/H evaluated at an age of 5 Gyr. With the best physics included, the mass for which the initial deuterium abundance is ultimately reduced by a factor 2 is found to be $13.3\,M_J$. If electron screening is neglected in the calculation of thermonuclear reaction rates, this mass rises to $13.6\,M_J$.

Despite major differences with the study of GG73, our mass limit for deuterium burning is very similar to theirs. But because we assume an initial deuterium abundance about one order of magnitude lower than the terrestrial value, we do not obtain a deuterium main sequence (defined to be a phase where $f_N = 1$).

We conclude that objects with masses below about 12 Jovian masses should retain essentially their entire initial complement of deuterium, and derive no luminosity at any stage in their evolution from thermonuclear fusion. Thus, the mass $12\,M_J$ represents a useful boundary to distinguish giant planets from brown dwarfs. Boss (1986) found that the minimum mass for protostars formed from collapse and fragmentation of Population I interstellar clouds was about $20\,M_J$. Lower-mass objects would have to form with the assistance of dense rock-ice cores, and thus would be considered giant planets. It is a coincidence that the limiting mass for giant planets defined by Boss' criterion and that defined by deuterium burning are nearly the same. However, we point out that the deuterium-burning criterion may prove to be the more useful of the two, since it can in principle be applied to observational data in an almost model-independent way. The calculation of the limiting mass for deuterium burning has proved to be very robust over two decades of improvement in the theory.

### 3.2. Inflation of Objects by the Central Star

As is well known (Hubbard 1977), the effect of photons from a primary star thermalized well below an EGP's photosphere is to modify the surface condition and direct the time evolution of the EGP toward an asymptotic effective temperature set only by the thermalized photons. An EGP with a companion star does not cool to zero temperature but tends toward an equilibrium temperature given by

$$T_{\mathrm{eq}} = \left[\frac{(1-A)L_\star}{16\pi\sigma a^2}\right]^{1/4}, \qquad (13)$$

where $A$ is the Bond albedo, $L_\star$ is the stellar luminosity, $\sigma$ is the Stefan-Boltzmann constant, and $a$ is the distance between the planet and the parent star.

Because thermal effects on the equation of state are significant for the lighter EGP's, the latter can also reach asymptotic radii significantly larger than that dictated by the zero-temperature equation of state. An EGP orbiting a luminous star will have a larger radius and higher luminosity than would be the case for an isolated EGP, which tends to offset the increased difficulty of detecting it against the greater background signal.

First, for isolated EGP's, Figs. 3 and 4, respectively, show surfaces of luminosity $L$ and radius $R$ as a function of time $t$ and mass $M$. These surfaces are terminated at a time $t = 5$ Gyr. The "ripple" in luminosity and radius at early times and for masses greater than $12\,M_J$ is caused by deuterium burning, as has been already discussed. Radii decline monotonically with time but show a more complicated behavior with mass. A radius minimum at early times transforms to a very broad maximum at about $4\,M_J$ for late times ($t > 1$ Gyr), and at late times radii are close to the mean radius of Jupiter, $70\,000$ km. We have plotted the observed values of $L$ and $R$ for Jupiter and Saturn in Figs. 3 and 4 respectively. These values differ somewhat from our models for solar-composition EGP's for two reasons. The observed values of $L$ for Jupiter and Saturn [$\log(L/L_\odot) = -9.062 \pm 0.034$ and $-9.651 \pm 0.030$ respectively] lie above the models at $t = 4.57$ Gyr because the models are calculated for isolated EGP's while thermalized photons from a G2V star are important for both Jupiter and Saturn (Conrath, Hanel & Samuelson 1989). And, in the case of Saturn, a further significant contribution to $L$ is possibly derived from immiscibility of helium and metallic-hydrogen mixtures (Salpeter 1973; Stevenson and Salpeter 1977a, 1977b; Hubbard and Stevenson 1984; Guillot et al. 1995), which is not included in our EGP models. Finally, it is well known that both Jupiter and Saturn possess dense cores composed of heavy elements, and that the $Z$-fraction of their mass is enhanced by roughly an order of magnitude over solar. It is this phenomenon, partly compensated by expansion due to rotation, that causes the mean radii of Jupiter and Saturn to plot about 3000 km and 6000 km, respectively, or about 4% and 10%, below the radii for solar-composition EGP's (Fig. 4). The sensitivity of $R$ and $L$ to several of the modeling assumptions is discussed in detail in §3.3.

To illustrate the pronounced effect of a luminous primary, Figs. 5 and 6 show the evolution of EGP's placed 10 AU from an A0V star. These surfaces are plotted on the same scale as Figs. 3 and 4, but are truncated at approximately the main sequence lifetime of the A0 star ($\sim 0.5$ Gyr). Behavior of the more massive EGP's is indistinguishable from Figs. 3 and 4, but for objects close to the mass of Jupiter and at late times, the effect of the thermalized photons is dominant and changes the evolution substantially. The lowest-mass EGP's ($\sim 1 M_J$) reach an asymptotic luminosity about 30 times higher than Jupiter's and their final radii stabilize at about 80 000 to 90 000 km. There seems to be no problem with the stability of such inflated objects against mass loss: the ratio of their radius to atmospheric scale height is always greater than 1000.

Our theory works quite well for Jupiter after allowance is made for modest inflation from a G2 star 5.2 AU distant (Burrows et al. 1995). In the case of Saturn, the larger discrepancies between our EGP model and observed parameters illustrate the increasing effect of nonsolar composition as well as possible immiscibility of hydrogen-helium mixtures. And, as Fig. 6 (b) makes clear, low-mass EGP's in the vicinity of luminous primaries will be greatly distended. In such an environment, EGP's with radii well in excess of 100 000 km may be found, particularly if they can form with masses below Saturn's mass ($M = 0.3 M_J$). However, the existence of such objects will depend upon whether they can form in an initially gravitationally-bound configuration.

### 3.3. Effects of rotation, helium abundance, and a dense core

The models presented assume a helium mass fraction of $Y = 0.25$, no rotation and no internal core consiting of heavy elements. On the other hand, we know that Jupiter and Saturn rotate rapidly, and possess a dense, central core which is probably formed from refractory materials from the protosolar nebula. Furthermore, we expect the composition of EGP's to differ from our assumed value of $Y = 0.25$. Using the method described in Guillot & Morel (1995) for the integration of the hydrostatic equilibrium in the presence of an "ice"+"rock" core, and assuming conservation of the angular momentum of the planet $\mathcal{M}_\omega$ during the evolution, we have investigated the influence of these parameters on the structure of EGP's of $1 M_J$ and $5 M_J$ located 5.2 A.U. from a G2V star. Our results are depicted on Figures 7 and 8, respectively. In these figures, we measure angular momentum in units of Jupiter's rotational angular momentum $\mathcal{M}_{\omega,J}$, and the core mass is in units of the Earth's mass $M_\oplus$.

As expected, the radius decreases as the abundance of helium $Y$ and the mass of the core $M_{\rm core}$, increase, and increases for a planet with a larger angular momentum. Quantitatively, the relative variations of the radius are almost independent of the age of the planet. However, the variations of the luminosity (excluding reflected starlight) shown in Figures 7 and 8 are more complex. Simple analytical models demonstrate that, without stellar insolation, the cooling of a hydrogen-helium object following an evolutionary path defined by $T_{10} \propto T_{\rm eff}^a$ with $a > 0$ is faster for larger radii (Hubbard 1977; Guillot et al. 1995). Inversely, for a given age, the luminosity *increases* with decreasing radius (see the variations of $L$ with $Y$ and $\mathcal{M}_\omega$ at 0.1 and 1 Gyr, in Figures 7 and 8). When the effect of stellar insolation becomes significant (at about 4.5 Gyr), the reverse can be true, i.e. an object with a larger radius will receive more energy from the parent star and can then be *more* luminous. On the other hand, Figures 7 and 8 show that the presence of a core tends to decrease the luminosity of the planet in spite of its slightly smaller radius. This is due to the reduced heat capacity of the planet.

For a realistic range of helium abundances, rotation rates, and core masses ($Y = 0.2 - 0.3$, $\mathcal{M}_\omega \leq \mathcal{M}_{\omega,J}$ for $1 M_J$ EGP's and $\mathcal{M}_\omega \leq 10 \mathcal{M}_{\omega,J}$ for $5 M_J$ EGP's, $M_{\rm core} < 20 M_\oplus$), we expect the variations of the radius and luminosity to be less than 5% at any age. Rapidly rotating EGP's cool faster and are more difficult to detect in the infrared, except when most of their luminosity is due to the energy absorbed from the parent star. Interestingly, planets with a larger abundance of helium are significantly brighter and therefore more easily detectable in the infrared (a more consistent calculation would include the effect of helium abundance on the atmospheric properties, but since the atmospheric absorption is not very sensitive to the helium abundance, this effect is small, at least for $Y < 0.4$). When observing the light *reflected* by EGP's in the visible, the opposite is true; with their smaller radius, planets with high $Y$ are more difficult to detect, whereas rapid rotators can be significantly brighter.

There are other effects which can cause departures from the assumed adiabatic interior profiles, such as a first-order phase transition, a radiative zone, the condensation of chemical species, or a phase separation. The effect of the presence of a first-order transition of molecular to metallic hydrogen has been investigated by Saumon et al. (1992) for Saturn, Jupiter and brown dwarfs and is found to be small. An increase of the intrinsic luminosity of about only a few percent is expected from the presence of the so-called plasma phase transition. The presence of a radiative region has more of an effect on planetary evolution. As studied for Saturn and Jupiter by Guillot et al. (1995), a radiative zone reduces the luminosity at a given age by about $15-20\%$ at a given age in Jupiter-like objects. More massive objects are expected to be less affected by this effect, as the increased absorption due to molecules like $H_2O$ or TiO at higher effective temperatures favors convective over radiative energy transport. The effect of condensation is twofold. First, it can lead to the presence of highly absorbing grains in the atmosphere and, therefore, change its properties. Second, it can affect the temperature profile, and then the internal structure of the planet. Unfortunately, these effects cannot be quantified without further studies. A phase separation of helium in hydrogen is postulated in Saturn in order to explain its high luminosity (Stevenson & Salpeter 1977b), and it is also possible in Jupiter (Guillot et al. 1995). Helium-hydrogen separation will yield smaller atmospheric helium abundances and higher luminosities. However, this occurs only in objects that are cold enough, i.e. for relatively low mass and old objects. The relative agreement between evolution models of Jupiter and the age of the solar system tells us that in planets of the size of Jupiter or larger, and at $t < 4.5\,\rm Gyr$, the relative increase of luminosity due to a possible phase separation of helium in hydrogen does not exceed 10%.

## 4. Spectral emission of extrasolar giant planets

To the best of our knowledge, there has been no work on the atmospheres and synthetic spectra of gaseous objects with effective temperatures of several hundred degrees which are typical of young and massive EGP's. The frequency dependence of the albedo and the phase function of such objects are therefore unknown, as well as the characteristics of the emitted spectra. While the giant planets of the solar system can be helpful guides at the low-$T_{\rm eff}$ limit of our calculation, the range demonstrated by their spectra serves as a warning against simple generalizations.

Given the trajectories for $L(M,t)$ and $R(M,t)$ presented in §3.2, approximate spectra for extrasolar giant planets can be generated easily. For lack of a better theory, we have assumed that the EGP's reflect the light of the parent star like a grey body and that the thermal emission is that of a blackbody. These approximations are adequate for the purpose of this calculation, which is to aid in designing search strategies and technological development for the detection of gas giants around nearby stars. A comparison with the actual spectrum of Jupiter is presented below.

The flux from an EGP is the sum of two separate contributions: intrinsic thermal emission (in the infrared) and reflected starlight (in the visible). Following the standard definitions (Mihalas 1978), the flux $\mathcal{F}_\nu$ received at the Earth from an EGP of radius $R$ orbiting at a distance $a$ from a star of radius $R_\star$ is given by:

$$\mathcal{F}_\nu = \left(\frac{R}{d}\right)^2 \mathcal{F}_\nu^{\rm p} + \frac{A}{4} P(\theta,\phi)\left(\frac{R_\star}{d}\right)^2 \left(\frac{R}{a}\right)^2 \mathcal{F}_\nu^\star \quad (14)$$

where $\mathcal{F}_\nu^\star$ is the flux radiated by the surface of the star, $\mathcal{F}_\nu^{\rm p} = \pi B_\nu(T_{\rm eff})$ is the thermal flux radiated by the surface of the planet and $d$ is the distance of the system from the Earth. The Bond albedo is $A$ and $P(\theta,\phi)$ is function which accounts for the angular dependence of the reflected light ($\int P(\theta,\phi)\,d\Omega = 4\pi$), where $\theta$ is the star-EGP-Earth angle. $P(\theta,\phi)$ can be measured for solar system objects or computed from the theory of planetary atmospheres. Here we assume the idealized case where the light reflected by the planet is redistributed uniformly over $4\pi$ steradian or $P(\theta, \phi) = 1$. While $P = 3.2$ for Jupiter in full phase ($\theta = 0$), we estimate that $P \approx 1$ for the quarter phase ($\theta = \pi/2$). This is the geometry which maximizes angular separation ($\theta = \pi/2$) between the EGP and its central star. If the global scattering properties of the atmospheres of EGP's do not differ dramatically from those of the giant planets of the solar system, our choice of $P = 1$ is representative of the most favorable phase for discovery.

In the next section, we estimate the anticipated deviations of the spectrum of EGP's from a black body spectrum. Section §4.2 presents the calculated fluxes from EGP's orbiting stars of spectral types A0V, G2V and M5V. The spectra $\mathcal{F}_\nu^\star$ of the A0V and G2V stars are obtained from the model for Vega by Dreiling & Bell (1980) and from the solar irradiance at Earth (A. Eibl, private communication), respectively. Since there are no measured spectra of M5 V stars which cover the wide wavelength range of interest, we use the $T_{\rm eff} = 3100\,{\rm K}$, $\log g = 5.5$ synthetic spectrum of Allard & Hauschildt (1995). These values of $T_{\rm eff}$ and $g$ are based on the main sequence models of Burrows et al. (1993) and the mass-spectral type relation derived by Kirkpatrick & McCarthy (1994). For stars of other spectral types, $\mathcal{F}_\nu^\star$ is taken from the synthetic spectra of Kurucz (1993). These spectra were extended into the far infrared with blackbody functions as needed.

Figure 9 displays the sensitivity of several ground and space-based observing platforms currently being developed and which will be applied to the search for extrasolar planets. These sensitivities are overlaid on Figs 10–13 and Fig. 15 where they can be directly compared with the predicted fluxes from EGP's. Except where noted below, the sensitivities plotted are for the detection of point sources with a signal-to-noise ratio of 5 in a one-hour integration. In all cases, except for the values at $0.8\,\mu{\rm m}$, the sensitivity is background-limited. The sensitivities of the Large Binocular Telescope (LBT) and the upgraded Multiple Mirror Telescope (MMT) at $0.8\,\mu{\rm m}$ are based on the application of adaptive optics with the high-order correction scheme proposed by Angel (1994). Diffraction-limited performance is expected from this new technology which however is limited to relatively bright stars ($R \leq 4$). The open triangles, 3-pointed stars and filled triangles show the sensitivities of cameras 1, 2, and 3, respectively, of the Near Infrared Camera and Multiple Object Spectrograph (NICMOS; G. Schneider, priv. comm.), a second generation instrument for the Hubble Space Telescope (HST). The integration time is limited to 40 minutes, as imposed by the HST orbit. Camera 2 has a $0.3''$ occulting disk to be used for planet searches. Thin solid lines show the sensitivity of the Infrared Space Observatory (ISO; ISOCAM Manual, 1994; ISOPHOT Manual, 1994). The length of the lines reflect the filter bandpasses. The Space InfraRed Telescope Facility (SIRTF) is the most sensitive mission currently in development at infrared wavelengths (shown by the thick solid lines; P. Eisenhardt, priv. comm.). Under the current design the resolution at the shortest wavelengths is $\sim 1-2''$. Finally, we also consider the Stratospheric Observatory For Infrared Astronomy (SOFIA) and the infrared capability of the Gemini telescope (dashed lines; P. Eisenhardt, priv. comm.). Except for ISO, which is built and scheduled for launch in 1995, all sensitivities given here represent the best current estimates at the present stage of development of each mission. A detailed comparison of predicted fluxes with instrumental sensitivities is deferred to §4.3.

### 4.1. Comparison of Jupiter's spectrum with the blackbody approximation

A comparison between our calculated spectrum for a $1\,M_{\rm J}$ EGP orbiting a G2V star at $5.2\,{\rm A.U.}$ (assuming blackbody emission) and the observed spectrum of Jupiter is presented in Fig. 10. Saturn has a very similar spectrum (e.g. Chamberlain & Hunten 1987; Karkoschka 1994), except that ammonia absorption is less intense because it condenses at deeper levels. For these planets, the flux can depart from blackbody emission by more than one order of magnitude at a given wavelength (*e.g.* at 2.2, 5, 6 $\mu$m). Almost all features in the Jovian spectrum correspond to molecular absorption bands. Because of the quasi-periodic structure in frequencies of the absorption of molecules such as $CH_4$, $NH_3$, and $H_2O$, these departures average out when using very broad spectroscopic bands (with extent larger than about $2000\,{\rm cm}^{-1}$). However, for practical reasons, observations will generally be constrained to narrower bands. It is therefore interesting to estimate where, for a given EGP, we expect the emitted flux to be higher than that of a blackbody.

The Jovian spectrum is, at large wavelengths ($\lambda \gtrsim 50\,\mu$m), dominated by strong absorption bands of $NH_3$. Similar absorption bands of $NH_3$ are present between 8 and 13 μm, 5 and 7 μm, 2.8 and 3.1 μm, 2.2 and 2.4 μm, 1.9 and 2.05 μm, and between 1.4 and 1.55 μm. The roto-translational band of collision-induced absorption by molecular hydrogen dominates from 13 μm to 45 μm, and the first vibrational band is centered on 2.5 μm. Methane has strong absorption bands centered around 7.5, 6, 3.5, 2.5, 1.7, 1.4, 1.15 and 1 μm. Water is buried deep in the Jovian atmosphere, so that it does not have a significant effect on the spectrum of the planet (though water lines are visible in the 5 μm region). For wavelengths between 0.4 and 1 μm, backscattering by ammonia clouds becomes very efficient, and the albedo of the planet approches 0.7 (compared to an average Bond albedo of 0.35). Narrow methane absorption bands are seen near 0.7, 0.9 and 1 μm. Clearly, the most favorable bands for observing Jupiter are between 0.4 and 1 μm, 4.5 and 5.3 μm and between 7 and 13 μm.

EGP's with effective temperatures below 150–200 K (corresponding to $1 M_J$ objects older than 0.4 Gyr, or 5 Gyr-old EGP's less massive than $5 M_J$, orbiting G2V or later stars) are expected to have spectra similar to that of Jupiter. For higher effective temperatures, the absorption by water clouds and water vapor is expected to play a dominant role in the spectrum. As the most common spectroscopic bands avoid the prominent absorption bands of water, observing in these bands should reveal both "cold" and "hot" EGP's with a comparable efficiency. Note however, that for hot enough objects, significant changes in the chemical composition (as the transformation of $CH_4$ to CO) are expected to yield major changes in their emission. Scattering by cloud particles is likely to increase the flux of all EGP's in the $B$, $V$, $R$ and $I$ bands. This effect should be even more significant when water condenses at low pressures (i.e. for "hot" EGP's), as seen on Venus, whose albedo is close to 0.9 in the $0.5 - 2.8$ μm region (e.g. Moroz 1983). On the other hand, methane (and possibly water) absorption bands strongly reduce the *reflected* flux in the 1–4 μm region, where no backwarming effect (e.g. Mihalas 1978) occurs, except for very hot EGP's. Hence, this spectroscopic region (which includes the $J$, $K$, and $H$ bands) is probably less favorable.

Thus, EGP's with effective temperatures between 200 and 300 K should be observed in the $B$, $V$, $R$, $I$, $L'$, $M$ and $N$ bands, where we expect the flux to be maximized. Colder EGP's ($T_{\text{eff}} \lesssim 200$ K) should be observed in the $B$, $V$, $R$, $I$, $M$ and $N$ bands, near 2.7 μm, and possibly at 1.25 μm ($J$ band), and 1.5 μm ($H$ band). EGP's in orbit around A0 stars always have effective temperatures above 300 K, even for a semi-major axis of 20 A.U. It is therefore very difficult to predict their emissions without a consistent thermochemical calculation. Nonetheless, it is probable that these objects are still dominated by absorption bands of water, and should therefore also be observed in the $B$, $V$, $R$, $I$, $L'$, $M$ and $N$ bands. Evidently, searches for EGP's should be conducted in several bandpasses.

### 4.2. Predicted fluxes

The flux received at the Earth from an EGP depends strongly on several parameters: the luminosity of the primary star $L_\star$, the semi-major axis of the orbit $a$, the mass of the planet $M_p$, its age $t$, and the distance of the system $d$. The combination of these parameters which occur in nature and that will lead to a successful detection of an EGP is not known *a priori*, but this rather broad parameter space can be somewhat constrained. Most searches are restricted to nearby stars and we adopt a representative distance of $d = 10$ pc. The nearby star sample has a median age of about 2 Gyr and it does not contain stars of spectral type earlier than A0. Furthermore, the example provided by the solar system, supplemented by protoplanetary disk models, limits the range of plausible values of $a$ within a few A.U. to a few tens of A.U.

In this section, we discuss a representative subset of our results. Combinations of primary stars and orbital radii that we consider in detail first include a system analogous to the solar system with a G2V star and EGP's orbiting at 5.2 A.U. (to allow a direct comparison with the familiar case of Jupiter) and at 10 A.U. There are 21 G dwarfs within 10 pc (T. Henry, priv. comm). Four A stars (Vega, Altair, Sirius, and Fomalhaut) occupy the bright end of the local population and we consider EGP's orbiting at 10 A.U. and 20 A.U. from an A0V star. Finally, the most numerous stars in the solar neighborhood are M dwarfs. As we will see, M dwarfs are so faint that their light contributes little if anything to the thermal emission of EGP's. However, in order to examine a case with significant flux of reflected light, we choose a smaller orbital radius of $a = 2.6$ A.U. Tables of fluxes in the $V$ through $N$ bandpasses for selected models are given in the Appendix.

The dependence of the predicted flux on the mass of EGP's in a system analogous to the Sun-Jupiter pair is shown in Fig. 10 where the planets are orbiting at 5.2 A.U. from a G2V star. The system is placed at 10 pc from the Sun and has the age of the solar system (4.5 Gyr). From left to right, the masses decrease from 14 $M_J$ to 0.3 $M_J$ (the mass of Saturn). The uppermost curve shows the spectrum of the star. Standard photometric bandpasses are indicated at the top of the figure and other symbols indicate the sensitivities of several instruments (see Fig. 9). After 4.5 Gyr of cooling, these planets have reached their final radius. For the masses considered, the radius of solar-composition planets falls in the narrow range of 1.01 to 1.09 $R_J$, where $R_J$ is Jupiter's radius. Except in cases of significant inflation (see §3.2), the radius of the EGP is essentially independent of its mass. It follows that the light reflected by the planet from primaries of type G2 and later is also independent of the mass. This is true in most cases because a broad maximum in the $R(M)$ relation is found around 4 $M_J$ for objects of solar composition and all EGP's therefore have similar radii. In general, the ratio of the luminosity reflected by the planet to that of the star is fixed by the separation ($\sim a^{-2}$) and is $\sim 3 \times 10^{-9}$ in Fig. 10.

On the other hand, the thermal emission of an EGP rises very rapidly with its mass in the 3 – 15 $\mu$m range, which corresponds to the Wein tail of the Planck function. In Figure 10, the analog of the Sun-Jupiter system is given by the 1 $M_J$ curve, which is the second from the right. Just doubling the mass to 2 $M_J$ can increase the flux by an order of magnitude at $N$, where the contrast between star and planet is reduced to $\sim 10^5 - 10^6$. At wavelengths above 30 $\mu$m, both the stellar and EGP spectra are in the Rayleigh-Jeans limit of the Planck function and the planet to star flux ratio becomes independent of wavelength:

$$\frac{\mathcal{F}_\nu^{\rm p}}{\mathcal{F}_\nu^\star} = \frac{T_{\rm eff}^{\rm p}}{T_{\rm eff}^\star}\frac{R^2}{R_\star^2}. \quad (15)$$

Various combinations of parameters involving a G2V central star 10 pc away from the Sun are shown in Figure 11. Panel $a$ show the evolution of the spectrum of a 1 $M_J$ EGP, orbiting at 5.2 A.U. The seven curves shown span ages from 0.01 to 5 Gyr, the rightmost curve corresponding approximately to the present Jupiter. Jupiter was much brighter in the past, with vigorous thermal emission. At 0.1 Gyr, Jupiter was as bright as a 14 $M_J$ EGP at the present age of the solar system (Fig. 10). On the other hand, the reflected light decreases very slowly with time, as EGP's at 5.2 A.U. from a G2V star are already within 30–40% of their final radius at 0.01 Gyr. The decrease in the reflected flux due to the contraction of the planet is less than a factor of two.

The spectral evolution of a 5 $M_J$ EGP is shown in Fig. 11b, where all other parameters are identical to those of panel $a$. As depicted in Fig. 3b, a 5 $M_J$ remains $\sim$ 10–20 times more luminous than a 1 $M_J$ at all times considered here. This translates into infrared fluxes which can be up to 1300 times larger than for a 1 $M_J$ in the $K$ through $N$ bands. For a given age, the highest flux ratios between these two masses are obtained in the band where the lower mass EGP emits only in reflected light while falling near, or blueward of the peak of thermal emission of the more massive EGP's. The emissions of a relatively old 5 $M_J$ EGP at 5 Gyr are nearly identical to those of a much younger 1 $M_J$ at the age of 0.3 Gyr.

Panel $c$ of Fig. 11 is similar to Fig. 10, but at $t = 1$ Gyr. Masses from 0.3 to 14 $M_J$ are shown. While all EGP's are brighter in this figure than in Fig. 10, there is hardly any change for the 0.3 $M_J$ (rightmost curve) since it reaches its final equilibrium temperature around 1.6 Gyr. On the other hand, at 1 Gyr the 14 $M_J$ model is still burning deuterium, albeit at a very slow rate. After 5 Gyr, all the deuterium has been consumed (as in Fig. 10).

If the orbital separation is increased, there is a corresponding decrease in the reflected flux ($\sim a^{-2}$) but the thermal flux is not affected until $T_{\rm eff}$ approaches $T_{\rm eq}$. The equilibrium temperature (Eq. 13) decreases with larger $a$, and the saturation effect on the spectrum occurs later. This is shown for a 0.3 $M_J$ EGP (Saturn's mass) in Fig. 11d, for $a = 5.2$ A.U. (solid lines) and $a = 10$ A.U. (dashed lines). The curves correspond to different ages ranging from 0.01 to 5 Gyr. For the first 0.1 Gyr of evolution, the thermal emission is dominated by internal heat, and stellar insolation is negligible. The spectrum of the EGP is independent of $a$ during this initial period. An EGP with as low a mass as 0.3 $M_J$ reaches its equilibrium configuration fairly early, in this case after 1.6 Gyr (for $a = 5.2$ A.U.). There is no further evolution of the EGP and its spectrum remains unchanged. This can be seen in the set of solid curves, where 3 Gyr and 5 Gyr are superposed. At a distance of 10 A.U., $T_{\rm eq}$ is lower and the EGP therefore cools for a longer time (dashed lines).

We illustrate the case of the brightest stars in the solar neighborhood by considering an A0V cen-tral star at 10 pc. Such a star has a mass of about 2.8 $M_\odot$, and a rather short main sequence lifetime of $\sim 0.4$ Gyr. Therefore, planetary systems around A stars are young. The four panels of Fig. 12 correspond to the panels of Fig. 11, with minor differences indicated below. Figures 12a and 12b show the evolution of 1 and 5 $M_J$ EGP's, respectively, orbiting at 10 A.U. from the star for ages of 0.01, 0.03, 0.1, 0.2, and 0.4 Gyr. The thermal emission during the early evolution is dominated by internal cooling and the absorbed heat from the star is negligible. This part of the spectrum is identical to Figs. 11a and 11b. On the other hand, an A0V star is $\sim 80$ times brighter than a G2V star and this results in a larger flux of reflected light. The steeper slope of the A0V spectrum also results in higher fluxes at shorter wavelengths.

The full range of masses considered (0.3 to 14 $M_J$) is shown in Fig. 12c for 0.2 Gyr, midway through the main sequence life of the A0V star. At this young age, the 14 $M_J$ EGP is still burning deuterium and emits vigorously in the near infrared. As discussed in §3.2, models of lower mass are significantly inflated by the absorbed stellar flux from the bright A star. As a consequence, the thermal emission of the 0.3 $M_J$ model is comparable to that of the 1 $M_J$ model.

The effect of increasing the planet–star separation from 10 A.U. (solid lines) to 20 A.U. (dashed lines) is shown in Fig. 12d. The spectra correspond to ages of 0.01, 0.03, 0.1, 0.2, and 0.4 Gyr for EGP's of 2 $M_J$. The effect on the thermal emission is not nearly as pronounced as in Fig. 11d because of the higher mass of the EGP considered here. Internal heat dominates the absorbed stellar flux in the more massive and younger objects. As expected, the reflected light is reduced by a factor of 4 as the distance from the central star is doubled to 20 A.U. This ratio would be larger for less massive EGP's which are inflated by absorbed stellar radiation.

Late M dwarfs represent a substantial pool of candidate stars for planetary system searches, since they are by far the most abundant type of stars in the solar neighborhood. Because they are intrinsically faint, the contrast between star and planet is minimized at infrared wavelengths for late M dwarfs. This is illustrated in Fig. 13 which considers EGP's orbiting a M5V star 10 pc away from the Sun. The luminosity of the star is so low ($L \approx 0.0034 L_\odot$) that it does not contribute to the energy balance of the EGP. Except for the reflected light, the spectrum of an EGP orbiting an M5V star is identical to that of an isolated EGP. To exhibit a case of significant reflected light flux, we adopt an orbital radius of 2.6 A.U. Using the semi-analytic protoplanetary disk model of Wood and Morfill (1988) and assuming that the disk accretion rate is independent of the mass of the central star and that Jupiter formed at the inner boundary of the zone where water ice condensed in the protosolar nebula, we can scale their disk model to other stars. The distance thus obtained for water ice condensation (2.6 A.U. for a M5V star) is more sensitive to the mass of the central star than predicted by the more detailed calculation of Boss (1995). The relatively small $a$ that we derive for a possible EGP companion to an M5V star gives some basis for searching such late stars for EGP emissions.

For $0.01 < t < 1$ Gyr, the evolution of the spectrum of a 1 $M_J$ EGP around a M5V star (Fig. 13a) is identical to that around a G2V star (Fig. 13a). The former will cool further to reach $T_{\rm eff} \approx 75$ K after a Hubble time, while the luminosity of the G2V primary will hold it at $T_{\rm eq} \approx 100$ K. Several curves are shown for the stellar spectrum in Fig. 13a,b, and d because an M5V star contracts for about 0.5 Gyr before settling on the main sequence. During this period, its luminosity decreases steadily at a nearly constant $T_{\rm eff}$. In reflected light, this EGP is one to three orders of magnitude fainter than its counterpart orbiting a G2V star. Because a 5 $M_J$ EGP is powered mostly by its internal heat during the period covered here ($t < 5$ Gyr), its thermal spectrum is identical to that of EGP's orbiting G2V and A0V stars. This is shown in Fig. 13b.

The spectra of EGP's with a range of masses (from 0.3 to 14 $M_J$) orbiting at 2.6 A.U. from a 1 Gyr old M5V star are shown in Fig 13c. Again, except for the reflected light, which is a few orders of magnitude fainter, the spectra are nearly identical to those of Fig. 11c. In the latter figure, the 0.3 $M_J$ model is twice as bright in thermal emission because of the light absorbed from the G2V star. Finally, since the light of the M5V star is so feeble, only the reflected light portion of the spectrum is affected by changing the orbital separation, $a$. This is shown in panel $d$ of Fig. 13 for a 2 $M_J$ EGP orbiting at 2.6 and 5.2 A.U. for ages of 0.01, 0.03, 0.1, 0.3, 1, 3, and 5 Gyr.

In the Rayleigh-Jeans limit of both the stellar and planetary spectra (in the far infrared), the flux from a 5 $M_J$ EGP is about 3% of that of the central star, a considerably lower contrast than would be the case with G2V or A0V primaries. Nevertheless, we do not predict any significant infrared excess in the energy distribution of any main sequence star due to the thermal emission of a giant planetary companion. Even in a most favorable case of a young system with a massive $10\,M_J$ planet, the flux ratio remains lower than 8%. Fig. 14 shows the flux ratio between a $2\,M_J$ EGP and a M5V star as a function of wavelength.

### 4.3. The potential for detection

As discussed in the previous section, the flux from EGP's represents only a small fraction of the flux of the parent star at any wavelength. Therefore, they are not detectable with photometric surveys unless the EGP and the star show substantial differences in their principal spectral features. Photometric and field imaging could in principle detect "free floating" EGP's, planet-sized bodies which are not bound to a star. This may be an unrealistic approach however; similar searches for the more massive and much brighter brown dwarfs have shown how difficult it is to find very dim objects, even in the solar neighborhood. In addition, it may also be impossible for such low-mass objects to form outside the environment of a dissipative protoplanetary accretion disk. It is more probable that planetary-mass objects will be found orbiting nearby stars, thereby qualifying as *bona fide* planets. The most compelling detection will come from resolving the EGP companion from its parent star by directly imaging the system.

Imaging giant planets around nearby stars presents major technological challenges. The difficulties arise mainly from the following problems: 1) The brightness ratio between the star and the planet is large and ranges from $\sim 30$ to $10^9$. 2) The small angular separation is perhaps $2''$ in a favorable case down to a more typical $0.5''$. 3) The flux from the planet is very weak. These three factors stretch the current limits of optical and infrared technologies. Current and next-generation instruments are reaching sensitivity levels within the range of the predicted brightnesses of EGP's, but they do not always have sufficiently high angular resolution to resolve the EGP from its parent star. The issue of angular resolution is further complicated by the problem of light scattered in the telescope optics. The point-spread function of diffraction-limited optical systems typically has a very faint halo which can spread over several arcseconds around the Airy disk, due to minute residual errors in the figure of the mirror, light scattered inside the telescope, or residual atmospheric distortions of the images. Because of the enormous contrast between the planet and the primary star, the signal of the planet can be lost in the halo of the primary star. The brightness of this faint halo is very difficult to predict and is expected to vary widely from one instrument to another.

Detection of planets by direct techniques, *i.e.* imaging using adaptive optics or interferometric techniques, must take into account the scattering of light by dust systems, analogous to our zodiacal light, around candidate stars. Further, such imaging in the mid-infrared is inhibited by our own zodiacal dust, requiring that infrared interferometers be placed in heliocentric orbits at 3 A.U. or beyond to avoid the worst of the dust emission (R. Angel, personal communication, 1995). Other star systems which are candidates for planetary searches may have higher dust column densities than our own, and hence more extreme scattering. Backman, et al. (1992) constructed dust density profiles for $\beta$ Pictoris from ground-based photometry in the visual and infrared. Their resulting profiles, even in the region expected to have planets (several to 30 A.U.) are several orders of magnitude higher than the solar system's zodiacal emission (C. Beichmann, pers. comm, 1995). Happily, the situation is much less severe if searches for giant planets are confined to expected separations of $\gtrsim 5$ A.U. from the parent star.

Although Saturn is the only planet with a ring system which is both radially broad ($\sim 10^5$ km) and optically thick, all giant planets of the solar system have rings and we expect them to occur in other planetary systems as well. With favorable geometry, an EGP with a ring system similar to that of Saturn could be a few times brighter in *reflected* light (visible wavelenghts). In the thermal infrared, the rings would reemit light absorbed from the central star at their equilibrium temperature which ranges from $\sim 30$ to 200 K for plausible systems. In most cases, this is smaller than the effective temperature of the planet and does not contribute much to the overall thermal flux.

The complexity and uncertainty which shroud these issues prevent us from providing a complete discussion of the detectability of EGP's. However, our calculations do predict their brightnesses, and the sensitivity of existing and projected instruments is reasonably well known. In the following discussion, we focus on the sensitivity of a set of representative instruments in the light of predicted EGP fluxes to comment on the detectability of EGP's in the near future. Other aspects of detection are brought into the discussion as deemed appropriate.

The three NICMOS cameras, with resolutions of 0.043, 0.075 and 0.2″ respectively, are very promising instruments for the detection of EGP's in the solar neighborhood. They are sensitive in the near infrared where most EGP's emit in reflected light. Figures 10–13 indicate that NICMOS will be able to detect EGP's around most types of central stars. Since the reflected light is nearly independent of the mass or the age of the planet, a wide variety of systems is within reach of this instrument. The high angular resolution afforded by HST should be amply adequate to resolve the systems shown in Figs. 10–13, which have an angular separation of 0.52″. The 0.3″ occulting disk of camera 2 should reduce the difficulties associated with the high contrast between the star and the planet. Extrasolar giant planets orbiting M5V stars are very faint in reflected light and NICMOS (as well as all other instruments currently in development) is not sensitive enough to be useful. However, Figure 13 shows that in the $J$, $H$, and $K$ bandpasses, it would pick up the thermal emission of the more massive and younger EGP's. A $1\,M_J$ EGP could be detected if only 10 millions years old (Fig. 13a). More massive EGP's stay bright longer and a $5\,M_J$ EGP could be seen at 10 pc for over 0.1 Gyr (Fig. 13b). However, these are rather optimistically young ages for M5V stars in the solar neighborhood. A more conservative value of 1 Gyr leads to the conclusion that only the most massive objects – $11\,M_J$ and above – could be seen by NICMOS at a distance of 10 pc (Fig. 13c).

With the adaptive optics scheme proposed by Angel (1994), both the MMT and the LBT will achieve diffraction-limited resolution ($\sim 0.025″$ and $\sim 0.014″$, respectively, at $0.8\,\mu$m) from the ground. Typical star/planet flux ratios which can be achieved for bright enough stars can be as high as $\sim 10^9$. Hence, the two telescopes will have sensitivities comparable to the NICMOS cameras at $\lambda = 0.8\,\mu$m (Fig. 9) and may successfully tackle the problem of scattered light. The requirement of a bright central star for accurate wavefront corrections excludes all M dwarfs in the solar neighborhood. The LBT sould be able to detect the reflected light of EGP's around all stars of earlier spectral types, regardless of the mass or the age of the planet, as long as the orbital radius $a$ is not too large (Fig. 11d). Resolving the planet and the star should be easy for any realistic $a$ within 10 pc. The MMT is about ten times less sensitive than the LBT and the combination of parameters that it can usefully search is limited to A-type stars (Fig. 12) or objects so young that they are unlikely to be found in the solar neighborhood. The sensitivity of the LBT in the $N$ band is limited by local thermal background and its potential for detecting the thermal emission of EGP's is considerably reduced at this wavelength. For EGP's orbiting stars of all spectral types (Figs 10 – 13) at a distance of 10 pc from the Sun, the LBT should see $1\,M_J$ planets when younger than 0.03 Gyr, and $5\,M_J$ objects would be detectable for up to 0.3 Gyr. A more reasonable age of 1 Gyr for a G2V star in the solar neighborhood limits the detection of EGP's to $M \gtrsim 12\,M_J$. However, the relative youth of A-type stars brings EGP's of masses above $5\,M_J$ within the range of detectability (Fig. 12c). The diffraction-limited resolution of the LBT is $\sim 0.18″$ at $N$, which is sufficient to resolve most systems within 10 pc.

EGP's of $1\,M_J$ which are bright enough to be seen in the mid- to far-infrared by Gemini and SOFIA are much too young to be found in the solar neighborhood. On the other hand, $5\,M_J$ EGP's are visible until they reach $\sim 0.5$ Gyr and by the time they are 1 Gyr old, planets more massive than $7\,M_J$ are detectable at 10 pc. The sensitivity of SOFIA is too low to be useful at wavelengths beyond $10\,\mu$m. The lower angular resolution of these telescopes ($\sim 1″$ at best at these wavelengths) limits useful searches to favorable systems with fairly large orbital radii and well within 10 pc of the Sun. It is not clear, however, whether Gemini and SOFIA will be able to achieve the proper level of contrast at small angular separations.

SIRTF will have the highest angular resolution of all space-based instruments in the mid- to far-infrared. Its high sensitivity gives it a real chance of detecting the thermal emission of EGP's in the solar neighborhood. It can detect $5\,M_J$ planets as old as 3 Gyr and for a somewhat younger system of 1 Gyr of age, $2\,M_J$ EGP's are accessible. Around an A0V star in particular, planets down to the mass of Saturn ($0.3\,M_J$) are detectable if orbiting within 10 A.U. At 20 A.U., the low-mass EGP's are not inflated as much by the absorbed stellar light and in this case the limit is $1\,M_J$. SIRTF should be particularly good at searching for EGP's around M dwarfs which are too faint in reflected light to be seen by other powerful instruments such as NICMOS and the LBT. Its expected angular resolution of $\sim 1-2″$ limits searches to favorable combinations of distance $d$ and orbital radius $a$.

We cannot presently address the issue of the contrast achievable with the SIRTF telescope and cameras.

The first EGP may well be discovered with ISO, which is scheduled for launch in 1995. Its 5 – 20 $\mu$m sensitivity is not much lower than that of SIRTF and it should see 3 $M_J$ planets around stars younger than 1 Gyr. Around the necessarily younger A0 stars, 0.3 $M_J$ planets are within reach. The reduced star/planet flux ratio obtained around M stars could result in a successful search for EGP's of 3 $M_J$ and above. The angular resolution of ISO is limited by its small aperture (0.6 m) and is further compromised by pointing jitter of 2.8″ (ISO Observer's Manual, 1994). This limits searches to stars within about 5 pc of the Sun. The potential for detection is real since this leaves ∼ 40 candidate stars to search.

We have selected four nearby stars as representative targets in search programs for EGP's: $\alpha$ Lyrae (Vega), $\tau$ Ceti, $\epsilon$ Eridani, and Gliese 699 (Barnard's star). In each case, we have calculated the evolution of 0.5, 1 and 3 $M_J$ EGP's with $a = 5.2$ A.U. Spectral type and parameters required for the calculation are given in Table 1 for each star. These stars were selected to cover the full range of main sequence spectral types found in the solar neighborhood and for the interest they have previously aroused regarding the possibility of planetary companions.

Vega is an A0V star with an infrared excess due to the presence of dust particles in orbit. Spectra for EGP's orbiting Vega are shown in Fig. 15a in which we adopt an age of 0.2 Gyr. An EGP companion to Vega could soon be detected. All observing platforms considered here are sensitive enough to detect EGP's of any mass around Vega. The angular dimension of a 5.2 A.U. orbit spans 0.67″ at Vega's distance of 7.72 pc which is below the angular resolution of ISO, SIRTF and SOFIA. Doubling the separation to 10 A.U. would bring it within reach of SIRTF. Instruments sensitive to reflected light (MMT, LBT and NICMOS) will easily resolve the EGP for any plausible separation.

Walker et al. (1995) have monitored $\tau$ Cet and $\epsilon$ Eri for radial velocity variations of very small amplitude during a 12-year period. While they obtained an indication of a possible companion to $\epsilon$ Eri with a period of about 10 years, they did not detect any companion of planetary mass around the 21 stars of their survey. Their analysis does not exclude EGP's of masses up to 2–3 $M_J$ with orbital periods from several years to over a decade. The shortest orbital period we anticipate for an EGP is based on the radius at which water ice condenses in the protoplanetary disk. Using the disk model of Wood & Morfill (1988) supplemented with the assumptions given in §4.2, we find that the shortest period should be ∼ 10 years. The disk model of Boss (1995) is consistent with this lower limit on the orbital periods of EGP's. The existence of EGP's with such long periods is only very weakly constrained by the 12-year study of Walker et al. (1995).

The target of several searches for extra-terrestrial intelligence, $\tau$ Ceti is a nearby solar-type star. Its main sequence lifetime is long and it shows no sign of chromospheric activity. Considering that the median age of stars in the solar neighborhood is 2 Gyr, we adopt a slightly older age of 3 Gyr for this calculation (Fig. 15b). The star is at 3.50 pc and a Jupiter-like separation would subtend a 1.5″ angle, too small for ISO to resolve. The MMT, LBT, and NICMOS will easily detect and resolve EGP's of all masses around $\tau$ Cet. SIRTF is sensitive enough to see a 3 $M_J$ planet beyond $\lambda = 8\,\mu$m. At this wavelength, a diffraction-limited resolution of only ∼ 2″ would require a separation slightly larger than 5.2 A.U. to resolve the system.

With a mass of only ∼ 0.8 $M_\odot$, $\epsilon$ Eridani has a very long main sequence life and could potentially be very old. However, its level of chromospheric activity and its far-infrared excess are signs of youth and we assign it an approximate age of 1 Gyr (Fig. 15c). EGP's orbiting $\epsilon$ Eri would therefore be brighter than around $\tau$ Cet. At 3.27 pc from the Sun, $\epsilon$ Eri is somewhat closer than $\tau$ Cet and the star/planet separation is 1.6″ for a 5.2 A.U. orbit. Again, reflected starlight will be easily detected by the MMT, the LBT and NICMOS. The younger age assumed for $\epsilon$ Eri brings 1 $M_J$ EGP's within reach of SIRTF and ISO. While ISO will not be able to resolve the system, SIRTF could do so if $a > 7$ A.U.

Barnard's star (Gl 699) was long thought to have an astrometric companion (van de Kamp 1986), but modern measurements have not confirmed the perturbations originally reported (Heintz 1994). Nevertheless, at 1.83 pc, it is the second nearest stellar system known (the closest being the $\alpha$ Cen triple system) and it is an interesting target. There is no indication of chromospheric activity in this star and we assume that it is fairly old at 3 Gyr (Fig. 15d). Barnard's star is too faint to be a viable target for the adaptive optics system planned for the MMT and LBT. The light reflected by the EGP is too feeble for the NIC-MOS cameras. SIRTF could easily see a $1\,M_J$ planet at 5.2 A.U. around Gl 699, since the corresponding angular separation is 2.84″. ISO could be equally successful if the system contained a $> 2\,M_J$ EGP and was slightly wider. Detectability is also favored by the greatly reduced contrast in the mid-infrared, which is $\sim 10^4 - 10^5$.

Both the Hyades and the Pleiades have been suggested as possible hunting grounds for extrasolar planetary systems. However, searches in these clusters may be in vain, at least in the near future. The Hyades form a fairly old cluster (0.6 Gyr) and are located 45 pc away from the Sun. The Pleiades, on the other hand, are nearly three times more distant at 126 pc but are much younger at 0.07 Gyr (but see Basri, Marcy & Graham 1995). Our calculation shows that the youth of the Pleiades far outweighs the larger distance since the bolometric luminosities of EGP's in the Pleiades would be 10 to 20 times larger than those of their cousins in the Hyades. In a few bandpasses, the flux can be up to two orders of magnitude larger. However, the relatively large distances of these two clusters impair EGP searches. In the Hyades, only the NICMOS cameras 1 and 2 have sufficient angular resolution to resolve a planet from its parent star (the stars are too faint to be viable targets for the MMT and LBT) and they will be able to see only very massive EGP's ($\gtrsim 12\,M_J$). The problem of resolution is even more acute in the more distant Pleiades, where only the NICMOS camera 1 will be able to resolve EGP's, and only if $a \gtrsim 8$ A.U. This instrument should reveal objects with $M > 7\,M_J$. Note that brown dwarfs, which are more massive, much brighter, and could form in isolation, should be easily detected by NICMOS, ISO, and SIRTF in these two clusters. The large arrays of the SIRTF cameras are very advantageous for such a program.

### 4.4. Detection of EGP's via transits of primaries

The radii of EGP's will generally lie in the range 80 000 to 100 000 km, depending on mass, age, and luminosity of the primary. An EGP with an age of 0.1 Gyr orbiting a main-sequence solar-type star will have a radius of approximately 90 000 km, or about 0.13 of its primary (Fig 4b). At the same time, its effective temperature will lie in the range 300 - 900 K (for a mass in the range $1 - 10\,M_J$), or about 0.05 – 0.15 of the primary's $T_{\rm eff}$. A transit of the primary by the EGP observed at optical wavelengths would thus lead to a maximum lightcurve depth of perhaps 1.6%. This is essentially the most favorable case for detecting EGP's orbiting solar-type stars in this fashion. But even for a Jupiter-class object at the present age of Jupiter, the maximum lightcurve depth remains at about 1.0%.

An EGP orbiting an A0V star at 10 A.U. will have a radius of about 90 000 km at 0.1 Gyr, although Saturn-mass EGP's could have radii in excess of 100 000 km. Taking the main-sequence radius of the A0 primary to be $1.1 \times 10^6$ km, the EGP would thus have a radius 0.08 of the primary, leading to a lightcurve depth of 0.6%, not greatly inferior to the situation for solar-type primaries.

In the case where an EGP orbits a very late main-sequence star with $M = 0.2\,M_\odot$ and $T_{\rm eff} = 3330$ K, a transit of the M dwarf by the EGP would lead to a high-amplitude lightcurve, for the EGP's radius would be in excess of 60% of the primary's.

We confirm the analysis of Borucki & Genet (1992), which shows that EGP's could be detected via transits of main-sequence primaries. An EGP will in general have a radius which is an appreciable fraction of the primary's radius, as is the case for Jupiter, and most EGP's will have radii somewhat larger than Jupiter's. In most cases transits could be reliably detected via photometry of the primary star at suitably-chosen wavelengths with a relative precision of $\sim 10^{-3}$.

### 5. Conclusions

We have constructed a broad suite of models of extrasolar giant planets, ranging in mass from 0.3 to $15\,M_J$. The models predict luminosity (both reflected and emitted) as a function of age, mass, deuterium abundance and distance from parent stars of various spectral type. We also explored the effect of variations in helium mass fraction and rotation rate and the effect of the presence of a rock-ice core.

The models employ the most accurate available equation of state for the interior, and boundary conditions interpolated between those of our previously published brown dwarf atmospheres (Burrows et al. 1993) and models optimized for low effective temperatures (GPGO). This enables us to predict with some confidence the radii of these objects as a function of mass and time, and to accurately characterize the inflation effect associated with illumination by the parent star.

Some of the primary conclusions of our study are

as follows:

1. Objects below 12 $M_J$ do not undergo deuterium fusion; we propose this limit as one way to distinguish brown dwarfs from EGPs (but we recognize that there are other distinguishing characteristics). For plausible values of the deuterium abundance in stars in the solar neighborhood, a deuterium main sequence ($f_N = 1$) is not obtained.

2. The interaction between illumination from the parent star and the radius of an EGP can lead to evolutionary histories distinctly different from those of isolated objects. Parent star illumination is primarily important for stars of solar and earlier spectral type. M dwarfs do not inflate EGP's at all for plausible orbital distances; except for reflected light, the behavior of EGP's around M-dwarfs is virtually identical to that of isolated EGP's.

3. The brightness contrast between an EGP and its parent star is a parameter of primary importance in gauging detectability. In the visible to near infrared this ratio varies primarily due to the surface area of the EGP and its distance from the parent star. Because the radius as a function of mass has a broad maximum around 4 $M_J$, the light reflected by EGP's is a very weak function of the mass. In contrast, the thermal emission and hence the brightness contrast in the 3–15 $\mu$m region of the spectrum varies sharply with mass: a doubling of the mass from 1 $M_J$ to 2 $M_J$ can decrease the star-planet contrast by an order of magnitude. Star-planet contrasts in the thermal infrared as low as 30 are found in some of our more massive EGP models, enormously more favorable than the value of $10^9$ found in the optical for a Jupiter orbiting a G2V star at 5 A.U. (a standard case for studies of planet searches).

4. The biggest challenge facing discovery of EGP's around nearby stars is the small angular separation (0.5–2″) between parent star and EGP, leading to the requirement for high effective angular resolution in direct detection strategies. The MMT and proposed LBT, configured for adaptive-optics systems, along with the air/spaceborne systems SOFIA, ISO, SIRTF and NICMOS on HST are collectively capable of detecting EGP's in the presence of parent star glare for much of the parameter space of wavelength, stellar spectral type, EGP mass and age explored here. Additionally, instruments designed to detect transits of stars by EGP's should do best for M dwarfs, provided they have sufficient photon sensitivity. Because each individual facility is capable of detection in a much smaller volume of this parameter space, we recommend that any planet search program utilize a variety of facilities over a broad range of wavelengths. Observations should be conducted in the $B$, $V$, $R$, $I$, $M$ and $N$ bands, where the emission of EGP's is expected to be maximized.

We close by emphasizing that the present set of models represents a significant improvement over previous studies in the mass range considered here. Black's (1980) pioneering models rely on the 12 $M_J$ model of GG73 to construct a power law relation for $L(M,t)$ and $R(M,t)$; because of the inaccuracy of this model, Black's relation overestimates luminosities by an order of magnitude and radii by 50% for 10 $M_J$-class models, becoming increasingly more accurate for less massive and older objects.

In spite of the careful treatment of boundary condition and interior physics, the present models represent only an intermediate step. Two major improvements required are (i) to treat the atmospheric thermal and reflected energy balance in a non-grey fashion, considering explicitly the frequency dependence of molecular absorptions and (ii) to model the properties of clouds for these atmospheres.

To understand the complexity associated with these problems, consider the evolution of an atmosphere as its effective temperature decreases from 1500 K to 150 K (Lunine et al. 1986). At roughly 1500 K silicate clouds condense and contribute opacity; these sink deeper into the interior as $T_{\text{eff}}$ drops. As $T_{\text{eff}}$ decreases below 1000 K, methane and additional water are formed at the expense of carbon monoxide, changing the atmospheric opacity. As the effective temperature drops below 200–300 K, molecular nitrogen converts to gaseous ammonia, and additionally water clouds condense around the one bar pressure level; both of these changes greatly affect the thermal and reflected opacity structures. Finally, the formation of ammonia clouds in the upper atmosphere at around Jupiter's effective temperature has a strong effect on reflected appearance.

Clearly, to model these changes accurately is a daunting challenge, but illustrative of the richness of EGP's as a class of astrophysical objects. Positive detections of such objects around nearby stars will intensify interest in such models; for now it is hoped the present comprehensive set of models will provide useful insights for the search.

We cannot present here all possible combinations of parameters of potential interest. A more complete set of tables is available on the Theoretical Astrophysics Program home page at the following address: (http://lepton.physics.arizona.edu:8000). Investigators who require further models should contact one of the authors to arrange for specific calculations.

We thank R. Angel, T. Henry, D. Hunten, J. Liebert, G. Rieke, D. Sandler, G. Schneider, A. Sprague, and N. Woolf for useful discussions. We are also grateful to G. Bjoraker for providing the IR spectrum of Jupiter shown in Fig. 10. This work was supported by NSF grant AST-9318970, NASA grant NAG5-2817 and by NASA grant HF-1051.01-93A from the Space Telescope Science Institute, which is operated by the Association of Universities for Research in Astronomy, Inc., under NASA contract NAS5-26555. T.G. acknowledges support from the European Space Agency.

## A. Tables of fluxes and apparent magnitudes

Tables A1–A5 give fluxes (in Janskys) received at the Earth for selected models. The fluxes are based on the approximations given in §4 and the transmission curves of Bessell & Brett (1988) and Bessell (1990) define the $V$ through $M$ photometric bandpasses. The $N$ bandpass is on the IRTF system (A. Tokunaga, priv. comm.).

---

This 2-column preprint was prepared with the AAS LATEX macros v4.0.

Fig. A1.— $T_{10}$ vs. $T_{\rm eff}$ for various surface gravities (in cgs units). For $T_{\rm eff} \leq 300\,{\rm K}$, surface conditions (3–4) are used (dots). The upper two curves are constrained by the gravity range of the atmosphere models calculated by GPGO, while the lower three curves represent extrapolations, via Eqs. (3–4), to higher gravities. For $T_{\rm eff} \geq 600\,{\rm K}$, the curves show the atmosphere models of the X sequence of Burrows et al. (1993).

Fig. A2.— (a) Luminosity vs. time for masses 10, 11, 12, 13, 14, and 15 $M_J$ (from bottom to top). The 14 $M_J$ model ignites deuterium later than the 15 $M_J$ model and thus has higher luminosity from about 0.1 to 1 Gyr. The heavier curves show the 13 $M_J$ model, a transition object which barely ignites deuterium. Solid curves show results when full (ion + electron) screening of the nuclear reaction is taken into account. Dotted curves show results when electron screening is neglected. The filled circle shows the low-luminosity end of the deuterium main sequence computed by GPGO (see text). (b) fraction of luminosity due to deuterium fusion, $f_N$, on the same time scale as panel (a). Because $f_N$ never reaches unity, there is no main sequence. (c) deuterium abundance as a function of mass for $\log t = 6.74, 7.11, 7.48, 7.85, 8.22$, and 9.70 (top to bottom).

Fig. A3.— (a) Surface of $L$ vs. $t$ and $M$ for isolated EGP's. The surface is terminated at $t = 5$ Gyr. Constant-mass contours for Jupiter and Saturn are highlighted, and observed values are plotted. (b) Projection of the surface onto the $L$ vs. $M$ plane, with isochrones for $\log t = 6.07, 6.37, 6.74, 7.11, 7.48, 7.85, 8.22, 8.59, 8.96, 9.33$, and 9.70 (top to bottom), along with observed values for Jupiter and Saturn.

Fig. A4.— (a) Surface of $R$ vs. $t$ and $M$ for isolated EGP's. The surface is terminated at $t = 5$ Gyr. Constant-mass contours for Jupiter and Saturn are highlighted, and observed values are plotted. The expansion of radii for low masses and early times is a consequence of the polytropic properties of fully-convective hydrogen-helium spheres at relatively high entropy. (b) Projection of the surface onto the $R$ vs. $M$ plane, with isochrones for $\log t = 6.07, 6.37, 6.74, 7.11, 7.48, 7.85, 8.22, 8.59, 8.96, 9.33$ and 9.70 (top to bottom), along with observed values for Jupiter and Saturn.

Fig. A5.— (a) Surface of $L$ vs. $t$ and $M$ for EGP's orbiting a main-sequence A0 star at 10 A.U. The surface is terminated at $t = 0.5$ Gyr. Observed values for Jupiter and Saturn are plotted. (b) Projection of the surface onto the $L$ vs. $M$ plane, with isochrones for $\log t = 6.05, 6.27, 6.53, 6.80, 7.07, 7.33, 7.60, 7.87, 8.14, 8.40$, and 8.67 (top to bottom), along with observed values for Jupiter and Saturn.

Fig. A6.— (a) Surface of $R$ vs. $t$ and $M$ for EGP's orbiting a main-sequence A0 star at 10 AU. The surface is terminated at $t = 0.5$ Gyr. Observed values for Jupiter and Saturn are plotted. (b) Projection of the surface onto the $R$ vs. $M$ plane, with isochrones for $\log t = 6.05, 6.27, 6.53, 6.80, 7.07, 7.33, 7.60, 7.87, 8.14, 8.40$, and 8.67 (top to bottom), along with observed values for Jupiter and Saturn.

Fig. A7.— Variations of the radius $R$ and total luminosity $L$ (excluding the reflected starlight) of a 1 $M_J$ (or $\sim 318\,M_\oplus$) EGP at 5.2 A.U. from a G2V star with the helium mass fraction $Y$, the angular momentum $\mathcal{M}_\omega$ (in units of the Jovian angular momentum $\mathcal{M}_{\omega,J}$), and the mass of a "rock"+"ice" core $M_{\rm core}$ (in units of Earth masses $M_\oplus$). The EGP's radius and luminosity are compared to a standard model with $Y = 0.25$, no rotation and no core, of radius $R_0$ and luminosity $L_0$. Different ages are represented: 0.1 Gyr (dotted lines), 1 Gyr (dashed lines), and 4.5 Gyr (solid lines). Note that an EGP with $Y = 1$ would be, at 4.5 Gyr, $\sim 2.5$ times brighter than our standard model (its *intrinsic* luminosity would be $\sim 6.4$ times larger). In models such that $\mathcal{M}_\omega > 3.5\mathcal{M}_{\omega,J}$, the centrifugal acceleration eventually becomes larger than the gravity during the contraction of the planet. Guillot et al. (1994) determine that the mass of the core is about 7 $M_\oplus$ in Jupiter, and between 0 and 20 $M_\oplus$ in Saturn.

Fig. A8.— Same as Figure 7, but for a 5 $M_J$ EGP. In this case, the insolation by the central star is less important, as the internal heat flux of the planet itself is much larger. With $Y = 1$, a 5 $M_J$ EGP is (after 4.5 Gyr), $\sim 2.8$ times brighter than the same EGP with $Y = 0.25$.

Fig. A9.— Sensitivities of ground-based and space-based observing platforms currently in development which will be applied to the search for extrasolar planets. The values quoted are for a 5-sigma detection of a point source in 1 hour of integration, except for the three NICMOS cameras, where the integration

Table A1
Characteristics of the nearby stars of Figure 15

| Star | Type | $L/L_\odot$ | $d$ (pc) | Age (Gyr) [a] |
|---|---|---|---|---|
| $\alpha$ Lyr | A0V | 80 | 7.72 | 0.2 |
| $\tau$ Cet | G8V | 0.59 | 3.50 | 3 |
| $\epsilon$ Eri | K2V | 0.34 | 3.27 | 1 |
| Gl 699 [b] | M4V | 0.0034 | 1.83 | 3 |

[a]Approximate age adopted for the calculation shown in Fig. 15.

[b]Barnard's star.

is limited to 40 minutes. The 0.8 µm sensitivities of the LBT and MMT depend on the brightness of the primary star. Standard photometric bandpasses are indicated at the top.

Fig. A10.— Spectral flux for EGP's orbiting at 5.2 A.U. from a G2V star at 10 pc from the Earth. Solid curves correspond to the following masses: 0.3, 1, 2, 4, 6, 8, 10, 12, and 14 $M_J$ (the more massive EGP being more luminous) at the age of the solar system: 4.5 Gyr or $\log t = 9.653$. A composite spectrum of Jupiter is superposed on the emission of the EGP's (Hanel et al. 1979; Hunten, Tomasko & Wallace 1980; Karkoschka 1994). The spectrum of the star is shown to scale. Other symbols are the same as in Fig. 9.

Fig. A11.— Spectral flux for EGP's orbiting a G2V star at 10 pc from the Earth. a) 1 $M_J$ EGP at 5.2 A.U. from the central star at ages of $\log t = 7$, 7.5, 8, 8.5, 9, 9.5, and 9.7 (from left to right). b) same as a, but for a 5 $M_J$ planet. c) EGP's of masses 0.3, 1, 2, 4, 6, 8, 10, 12, and 14 $M_J$ (from right to left) orbiting at 5.2 A.U. and at an age of $\log t = 9$. d) 0.3 $M_J$ planet orbiting at 5.2 A.U. (solid lines) and 10 A.U. (dashed lines) at times of $\log t = 7$, 7.5, 8, 8.5, 9, 9.5, and 9.7 (from left to right). The spectrum of the star is shown to scale. Other symbols are the same as in Fig. 9.

Fig. A12.— Spectral flux for EGP's orbiting an A0V star at 10 pc from the Earth. a) 1 $M_J$ EGP at 10 A.U. from the central star at ages of $\log t = 7$, 7.5, 8, 8.3, and 8.6. b) same as a, but for a 5 $M_J$ planet. c) EGP's of masses 0.3, 1, 2, 4, 6, 8, 10, 12, and 14 $M_J$ orbiting at 10 A.U. and at an age of $\log t = 8.3$. d) 2 $M_J$ planet orbiting at 10 A.U. (solid lines) and 20 A.U. (dashed lines) at times of $\log t = 7$, 7.5, 8, 8.3, and 8.6. The spectrum of the star is shown to scale. Other symbols are the same as in Fig. 9.

Fig. A13.— Spectral flux for EGP's orbiting a M5V star at 10 pc from the Earth. a) 1 $M_J$ EGP at 2.6 A.U. from the central star at ages of $\log t = 7$, 7.5, 8, 8.5, 9, 9.5, and 9.7 (from left to right). b) same as a, but for a 5 $M_J$ planet. c) EGP's of masses 0.3, 1, 2, 4, 6, 8, 10, 12, and 14 $M_J$ (from right to left) orbiting at 2.6 A.U. and at an age of $\log t = 9$. d) 2 $M_J$ planet orbiting at 2.6 A.U. (solid lines) and 5.2 A.U. (dashed lines) at times of $\log t = 7$, 7.5, 8, 8.5, 9, 9.5, and 9.7 (from left to right). The spectrum of the star is shown to scale at the top and the luminosity of the M5V star decreases as it contracts towards the main sequence. Contraction stops when the ZAMS is reached shortly before $\log t = 9$. Other symbols are the same as in Fig. 9.

Fig. A14.— Flux ratio between the planet and the primary star for a 2 $M_J$ planet orbiting 2.6 A.U. from a M5V star. Curves, from left to right, are for $\log t = 7$, 7.5, 8, 8.5, 9, 9.5, and 9.7. Standard photometric bandpasses are shown at the top.

Fig. A15.— Spectral flux from hypothetical EGP's orbiting at $a = 5.2$ A.U. from specific nearby stars (see Table 1). The spectra correspond to EGP's of 0.5, 1 and 3 $M_J$ (from right to left) . a) $\alpha$ Lyr (Vega), b) $\tau$ Cet, c) $\epsilon$ Eri, and d) Gl 699 (Barnard's star). The stellar spectra are shown to scale. The far-infrared excesses of $\alpha$ Lyr and $\epsilon$ Eri observed by IRAS are shown by dashed lines (Gillett 1986). Other symbols are the same as in Fig. 9.

TABLE A1
Planet orbiting at 5.2 AU from a G2V star at 10.0 pc from Earth

| | | | | $\log \mathcal{F}_\nu$ (Jy) | | | | | |
|---|---|---|---|---|---|---|---|---|---|
| $M/M_J$ | $\log t$ | V | R | I | J | H | K | M | N |
| 0.299 | 7.50 | −6.65 | −6.56 | −6.52 | −6.52 | −6.59 | −6.80 | −6.57 | −4.63 |
|  | 8.00 | −6.74 | −6.66 | −6.62 | −6.62 | −6.68 | −6.89 | −7.38 | −5.42 |
|  | 8.50 | −6.82 | −6.73 | −6.69 | −6.69 | −6.76 | −6.97 | −7.51 | −6.24 |
|  | 9.00 | −6.87 | −6.79 | −6.74 | −6.74 | −6.81 | −7.02 | −7.57 | −7.08 |
|  | 9.50 | −6.89 | −6.80 | −6.76 | −6.76 | −6.82 | −7.04 | −7.58 | −7.43 |
|  | 9.70 | −6.89 | −6.80 | −6.76 | −6.76 | −6.83 | −7.04 | −7.58 | −7.45 |
| 0.500 | 7.50 | −6.65 | −6.57 | −6.53 | −6.53 | −6.59 | −6.81 | −5.65 | −4.17 |
|  | 8.00 | −6.73 | −6.64 | −6.60 | −6.60 | −6.66 | −6.88 | −6.95 | −4.88 |
|  | 8.50 | −6.79 | −6.70 | −6.66 | −6.66 | −6.72 | −6.94 | −7.46 | −5.69 |
|  | 9.00 | −6.84 | −6.75 | −6.71 | −6.71 | −6.78 | −6.99 | −7.54 | −6.59 |
|  | 9.50 | −6.87 | −6.78 | −6.74 | −6.74 | −6.81 | −7.02 | −7.56 | −7.28 |
|  | 9.70 | −6.87 | −6.79 | −6.75 | −6.75 | −6.81 | −7.02 | −7.57 | −7.39 |
| 0.700 | 7.50 | −6.66 | −6.58 | −6.54 | −6.54 | −6.60 | −6.81 | −5.01 | −3.87 |
|  | 8.00 | −6.72 | −6.64 | −6.60 | −6.60 | −6.66 | −6.87 | −6.28 | −4.51 |
|  | 8.50 | −6.77 | −6.69 | −6.65 | −6.65 | −6.71 | −6.93 | −7.35 | −5.30 |
|  | 9.00 | −6.82 | −6.74 | −6.70 | −6.70 | −6.76 | −6.97 | −7.52 | −6.20 |
|  | 9.50 | −6.85 | −6.77 | −6.73 | −6.73 | −6.79 | −7.01 | −7.55 | −7.06 |
|  | 9.70 | −6.86 | −6.78 | −6.74 | −6.74 | −6.80 | −7.01 | −7.56 | −7.28 |
| 1.000 | 7.50 | −6.68 | −6.59 | −6.55 | −6.55 | −6.61 | −6.69 | −4.38 | −3.57 |
|  | 8.00 | −6.73 | −6.65 | −6.60 | −6.60 | −6.67 | −6.88 | −5.61 | −4.19 |
|  | 8.50 | −6.77 | −6.69 | −6.65 | −6.65 | −6.71 | −6.92 | −7.02 | −4.95 |
|  | 9.00 | −6.81 | −6.73 | −6.69 | −6.69 | −6.75 | −6.96 | −7.49 | −5.78 |
|  | 9.50 | −6.84 | −6.76 | −6.72 | −6.72 | −6.78 | −7.00 | −7.54 | −6.71 |
|  | 9.70 | −6.85 | −6.77 | −6.73 | −6.73 | −6.79 | −7.00 | −7.55 | −7.03 |
| 1.200 | 7.50 | −6.68 | −6.60 | −6.56 | −6.56 | −6.61 | −6.45 | −4.11 | −3.44 |
|  | 8.00 | −6.73 | −6.65 | −6.61 | −6.61 | −6.67 | −6.88 | −5.28 | −4.03 |
|  | 8.50 | −6.77 | −6.69 | −6.64 | −6.64 | −6.71 | −6.92 | −6.75 | −4.78 |
|  | 9.00 | −6.81 | −6.72 | −6.68 | −6.68 | −6.74 | −6.96 | −7.47 | −5.58 |
|  | 9.50 | −6.84 | −6.75 | −6.71 | −6.71 | −6.77 | −6.99 | −7.53 | −6.50 |
|  | 9.70 | −6.85 | −6.76 | −6.72 | −6.72 | −6.78 | −7.00 | −7.54 | −6.85 |
| 1.500 | 7.50 | −6.68 | −6.60 | −6.56 | −6.56 | −6.59 | −6.02 | −3.84 | −3.30 |
|  | 8.00 | −6.73 | −6.65 | −6.60 | −6.60 | −6.67 | −6.87 | −4.92 | −3.86 |
|  | 8.50 | −6.77 | −6.68 | −6.64 | −6.64 | −6.70 | −6.92 | −6.38 | −4.58 |
|  | 9.00 | −6.80 | −6.72 | −6.67 | −6.68 | −6.74 | −6.95 | −7.38 | −5.32 |
|  | 9.50 | −6.83 | −6.75 | −6.70 | −6.71 | −6.77 | −6.98 | −7.53 | −6.23 |
|  | 9.70 | −6.84 | −6.76 | −6.71 | −6.72 | −6.78 | −6.99 | −7.54 | −6.59 |
| 2.000 | 7.50 | −6.68 | −6.60 | −6.56 | −6.55 | −6.42 | −5.39 | −3.51 | −3.14 |
|  | 8.00 | −6.73 | −6.64 | −6.60 | −6.60 | −6.66 | −6.77 | −4.49 | −3.65 |
|  | 8.50 | −6.76 | −6.68 | −6.64 | −6.64 | −6.70 | −6.91 | −5.82 | −4.31 |
|  | 9.00 | −6.79 | −6.71 | −6.67 | −6.67 | −6.73 | −6.94 | −7.10 | −5.01 |
|  | 9.50 | −6.82 | −6.74 | −6.70 | −6.70 | −6.76 | −6.97 | −7.51 | −5.87 |
|  | 9.70 | −6.83 | −6.75 | −6.71 | −6.71 | −6.77 | −6.98 | −7.53 | −6.24 |
| 3.000 | 7.50 | −6.68 | −6.59 | −6.55 | −6.48 | −5.64 | −4.54 | −3.10 | −2.92 |
|  | 8.00 | −6.72 | −6.64 | −6.59 | −6.59 | −6.64 | −6.17 | −3.93 | −3.37 |
|  | 8.50 | −6.76 | −6.67 | −6.63 | −6.63 | −6.69 | −6.90 | −5.07 | −3.94 |
|  | 9.00 | −6.78 | −6.70 | −6.66 | −6.66 | −6.72 | −6.94 | −6.41 | −4.61 |
|  | 9.50 | −6.81 | −6.73 | −6.69 | −6.69 | −6.75 | −6.96 | −7.42 | −5.39 |
|  | 9.70 | −6.82 | −6.74 | −6.70 | −6.70 | −6.76 | −6.98 | −7.50 | −5.72 |
| 5.000 | 7.50 | −6.67 | −6.58 | −6.54 | −5.53 | −4.39 | −3.56 | −2.63 | −2.66 |
|  | 8.00 | −6.72 | −6.63 | −6.59 | −6.57 | −6.06 | −4.92 | −3.30 | −3.05 |
|  | 8.50 | −6.76 | −6.67 | −6.63 | −6.63 | −6.69 | −6.55 | −4.21 | −3.53 |
|  | 9.00 | −6.79 | −6.70 | −6.66 | −6.66 | −6.72 | −6.94 | −5.48 | −4.16 |
|  | 9.50 | −6.81 | −6.73 | −6.69 | −6.69 | −6.75 | −6.96 | −6.71 | −4.77 |
|  | 9.70 | −6.82 | −6.74 | −6.70 | −6.70 | −6.76 | −6.97 | −7.20 | −5.10 |
| 10.000 | 7.50 | −6.64 | −6.16 | −5.33 | −3.61 | −2.90 | −2.44 | −2.09 | −2.34 |
|  | 8.00 | −6.74 | −6.65 | −6.57 | −5.07 | −4.03 | −3.31 | −2.55 | −2.65 |
|  | 8.50 | −6.79 | −6.71 | −6.67 | −6.56 | −5.65 | −4.56 | −3.17 | −3.02 |
|  | 9.00 | −6.83 | −6.75 | −6.71 | −6.71 | −6.76 | −6.47 | −4.17 | −3.54 |
|  | 9.50 | −6.85 | −6.77 | −6.73 | −6.73 | −6.79 | −7.00 | −5.63 | −4.26 |
|  | 9.70 | −6.86 | −6.77 | −6.73 | −6.73 | −6.80 | −7.01 | −6.00 | −4.44 |

TABLE A2
Planet orbiting at 10.0 AU from a G2V star at 10.0 pc from Earth

| $M/M_J$ | $\log t$ | V | R | I | J | H | K | M | N |
|---|---|---|---|---|---|---|---|---|---|
| | | | | $\log \mathcal{F}_\nu$ (Jy) | | | | | |
| 0.299 | 7.50 | −7.22 | −7.13 | −7.09 | −7.09 | −7.16 | −7.37 | −6.66 | −4.64 |
| | 8.00 | −7.31 | −7.23 | −7.19 | −7.19 | −7.25 | −7.47 | −7.86 | −5.46 |
| | 8.50 | −7.40 | −7.31 | −7.27 | −7.27 | −7.33 | −7.55 | −8.09 | −6.41 |
| | 9.00 | −7.46 | −7.37 | −7.33 | −7.33 | −7.40 | −7.61 | −8.16 | −7.60 |
| | 9.50 | −7.50 | −7.41 | −7.37 | −7.37 | −7.44 | −7.65 | −8.19 | −8.48 |
| | 9.70 | −7.51 | −7.42 | −7.38 | −7.38 | −7.45 | −7.66 | −8.20 | −8.62 |
| 0.500 | 7.50 | −7.22 | −7.14 | −7.10 | −7.10 | −7.16 | −7.37 | −5.66 | −4.17 |
| | 8.00 | −7.29 | −7.21 | −7.17 | −7.17 | −7.23 | −7.45 | −7.10 | −4.90 |
| | 8.50 | −7.36 | −7.27 | −7.23 | −7.23 | −7.29 | −7.51 | −8.00 | −5.75 |
| | 9.00 | −7.42 | −7.33 | −7.29 | −7.29 | −7.35 | −7.57 | −8.11 | −6.84 |
| | 9.50 | −7.46 | −7.37 | −7.33 | −7.33 | −7.40 | −7.61 | −8.15 | −8.04 |
| | 9.70 | −7.47 | −7.39 | −7.34 | −7.34 | −7.41 | −7.62 | −8.17 | −8.36 |
| 1.000 | 7.50 | −7.25 | −7.16 | −7.12 | −7.12 | −7.18 | −6.99 | −4.36 | −3.56 |
| | 8.00 | −7.30 | −7.21 | −7.17 | −7.17 | −7.23 | −7.45 | −5.61 | −4.19 |
| | 8.50 | −7.34 | −7.26 | −7.21 | −7.22 | −7.28 | −7.49 | −7.18 | −4.96 |
| | 9.00 | −7.38 | −7.30 | −7.26 | −7.26 | −7.32 | −7.53 | −8.05 | −5.87 |
| | 9.50 | −7.42 | −7.33 | −7.29 | −7.29 | −7.35 | −7.57 | −8.11 | −7.01 |
| | 9.70 | −7.43 | −7.35 | −7.30 | −7.30 | −7.37 | −7.58 | −8.13 | −7.53 |
| 2.000 | 7.50 | −7.25 | −7.16 | −7.12 | −7.11 | −6.69 | −5.41 | −3.51 | −3.14 |
| | 8.00 | −7.29 | −7.21 | −7.17 | −7.17 | −7.23 | −7.14 | −4.49 | −3.65 |
| | 8.50 | −7.33 | −7.25 | −7.20 | −7.20 | −7.27 | −7.48 | −5.84 | −4.31 |
| | 9.00 | −7.36 | −7.28 | −7.24 | −7.24 | −7.30 | −7.51 | −7.27 | −5.02 |
| | 9.50 | −7.39 | −7.31 | −7.27 | −7.27 | −7.33 | −7.54 | −8.07 | −5.95 |
| | 9.70 | −7.40 | −7.32 | −7.28 | −7.28 | −7.34 | −7.56 | −8.10 | −6.38 |
| 3.000 | 7.50 | −7.24 | −7.16 | −7.12 | −6.89 | −5.68 | −4.54 | −3.10 | −2.92 |
| | 8.00 | −7.29 | −7.20 | −7.16 | −7.16 | −7.15 | −6.24 | −3.94 | −3.37 |
| | 8.50 | −7.32 | −7.24 | −7.20 | −7.20 | −7.26 | −7.45 | −5.08 | −3.95 |
| | 9.00 | −7.35 | −7.27 | −7.23 | −7.23 | −7.29 | −7.50 | −6.46 | −4.61 |
| | 9.50 | −7.38 | −7.30 | −7.26 | −7.26 | −7.32 | −7.53 | −7.84 | −5.43 |
| | 9.70 | −7.39 | −7.31 | −7.27 | −7.27 | −7.33 | −7.54 | −8.04 | −5.78 |
| 4.000 | 7.50 | −7.24 | −7.15 | −7.11 | −6.23 | −4.94 | −3.97 | −2.83 | −2.77 |
| | 8.00 | −7.29 | −7.20 | −7.16 | −7.15 | −6.76 | −5.48 | −3.57 | −3.19 |
| | 8.50 | −7.32 | −7.24 | −7.20 | −7.20 | −7.26 | −7.22 | −4.58 | −3.71 |
| | 9.00 | −7.35 | −7.27 | −7.22 | −7.23 | −7.29 | −7.50 | −5.89 | −4.35 |
| | 9.50 | −7.38 | −7.29 | −7.25 | −7.25 | −7.32 | −7.53 | −7.33 | −5.06 |
| | 9.70 | −7.39 | −7.30 | −7.26 | −7.26 | −7.33 | −7.54 | −7.82 | −5.40 |
| 6.000 | 7.50 | −7.24 | −7.14 | −6.96 | −5.03 | −3.98 | −3.25 | −2.48 | −2.58 |
| | 8.00 | −7.29 | −7.20 | −7.16 | −6.86 | −5.61 | −4.49 | −3.10 | −2.94 |
| | 8.50 | −7.33 | −7.24 | −7.20 | −7.20 | −7.15 | −6.12 | −3.90 | −3.37 |
| | 9.00 | −7.36 | −7.28 | −7.24 | −7.24 | −7.30 | −7.49 | −5.13 | −3.99 |
| | 9.50 | −7.39 | −7.30 | −7.26 | −7.26 | −7.32 | −7.54 | −6.44 | −4.62 |
| | 9.70 | −7.40 | −7.31 | −7.27 | −7.27 | −7.33 | −7.55 | −7.04 | −4.92 |
| 8.000 | 7.50 | −7.23 | −6.93 | −6.17 | −4.18 | −3.34 | −2.77 | −2.25 | −2.44 |
| | 8.00 | −7.30 | −7.21 | −7.16 | −5.88 | −4.66 | −3.78 | −2.76 | −2.76 |
| | 8.50 | −7.35 | −7.26 | −7.22 | −7.19 | −6.48 | −5.21 | −3.47 | −3.16 |
| | 9.00 | −7.38 | −7.30 | −7.26 | −7.26 | −7.32 | −7.20 | −4.56 | −3.73 |
| | 9.50 | −7.40 | −7.32 | −7.28 | −7.28 | −7.34 | −7.56 | −6.00 | −4.43 |
| | 9.70 | −7.41 | −7.33 | −7.29 | −7.29 | −7.35 | −7.56 | −6.44 | −4.64 |
| 10.000 | 7.50 | −7.13 | −6.30 | −5.35 | −3.61 | −2.90 | −2.44 | −2.09 | −2.34 |
| | 8.00 | −7.31 | −7.21 | −7.02 | −5.08 | −4.03 | −3.31 | −2.55 | −2.65 |
| | 8.50 | −7.36 | −7.28 | −7.24 | −6.93 | −5.68 | −4.57 | −3.17 | −3.02 |
| | 9.00 | −7.40 | −7.31 | −7.27 | −7.27 | −7.30 | −6.58 | −4.17 | −3.54 |
| | 9.50 | −7.42 | −7.34 | −7.30 | −7.30 | −7.36 | −7.57 | −5.64 | −4.27 |
| | 9.70 | −7.43 | −7.34 | −7.30 | −7.30 | −7.36 | −7.58 | −6.02 | −4.45 |

TABLE A3
Planet orbiting at 10.0 AU from a A0V star at 10.0 pc from Earth

| | | $\log \mathcal{F}_\nu$ (Jy) | | | | | | | |
|---|---|---|---|---|---|---|---|---|---|
| $M/M_J$ | $\log t$ | V | R | I | J | H | K | M | N |
| 0.299 | 7.00 | −5.43 | −5.50 | −5.60 | −5.78 | −5.97 | −6.17 | −5.08 | −3.83 |
| | 7.50 | −5.52 | −5.59 | −5.69 | −5.87 | −6.07 | −6.27 | −5.91 | −4.28 |
| | 8.00 | −5.56 | −5.63 | −5.73 | −5.91 | −6.10 | −6.30 | −6.27 | −4.49 |
| | 8.30 | −5.56 | −5.64 | −5.74 | −5.92 | −6.11 | −6.31 | −6.31 | −4.52 |
| 0.500 | 7.00 | −5.49 | −5.56 | −5.66 | −5.84 | −6.03 | −6.17 | −4.37 | −3.52 |
| | 7.50 | −5.56 | −5.64 | −5.74 | −5.92 | −6.11 | −6.31 | −5.35 | −4.03 |
| | 8.00 | −5.61 | −5.69 | −5.79 | −5.97 | −6.16 | −6.36 | −6.11 | −4.43 |
| | 8.30 | −5.63 | −5.70 | −5.80 | −5.98 | −6.17 | −6.37 | −6.32 | −4.55 |
| 1.000 | 7.00 | −5.54 | −5.61 | −5.71 | −5.89 | −5.85 | −5.12 | −3.42 | −3.07 |
| | 7.50 | −5.60 | −5.68 | −5.78 | −5.96 | −6.14 | −6.22 | −4.29 | −3.54 |
| | 8.00 | −5.65 | −5.72 | −5.82 | −6.00 | −6.19 | −6.39 | −5.33 | −4.07 |
| | 8.30 | −5.67 | −5.74 | −5.84 | −6.02 | −6.21 | −6.42 | −5.91 | −4.35 |
| 2.000 | 7.00 | −5.55 | −5.62 | −5.72 | −5.65 | −4.65 | −3.88 | −2.79 | −2.74 |
| | 7.50 | −5.61 | −5.68 | −5.78 | −5.96 | −5.92 | −5.19 | −3.49 | −3.14 |
| | 8.00 | −5.65 | −5.73 | −5.83 | −6.01 | −6.20 | −6.30 | −4.42 | −3.63 |
| | 8.30 | −5.67 | −5.75 | −5.85 | −6.03 | −6.22 | −6.41 | −5.05 | −3.95 |
| 3.000 | 7.00 | −5.55 | −5.62 | −5.72 | −4.94 | −3.91 | −3.27 | −2.48 | −2.56 |
| | 7.50 | −5.60 | −5.68 | −5.78 | −5.91 | −5.26 | −4.42 | −3.09 | −2.92 |
| | 8.00 | −5.65 | −5.72 | −5.82 | −6.00 | −6.16 | −5.85 | −3.90 | −3.37 |
| | 8.30 | −5.67 | −5.74 | −5.84 | −6.02 | −6.21 | −6.34 | −4.49 | −3.67 |
| 4.000 | 7.00 | −5.54 | −5.61 | −5.67 | −4.31 | −3.41 | −2.87 | −2.28 | −2.44 |
| | 7.50 | −5.60 | −5.67 | −5.77 | −5.67 | −4.66 | −3.90 | −2.82 | −2.77 |
| | 8.00 | −5.65 | −5.72 | −5.82 | −6.00 | −5.98 | −5.26 | −3.55 | −3.18 |
| | 8.30 | −5.67 | −5.74 | −5.84 | −6.02 | −6.20 | −6.07 | −4.09 | −3.47 |
| 6.000 | 7.00 | −5.53 | −5.52 | −5.12 | −3.45 | −2.76 | −2.35 | −2.01 | −2.27 |
| | 7.50 | −5.60 | −5.67 | −5.76 | −4.82 | −3.82 | −3.21 | −2.48 | −2.58 |
| | 8.00 | −5.65 | −5.72 | −5.82 | −5.93 | −5.22 | −4.38 | −3.09 | −2.94 |
| | 8.30 | −5.67 | −5.75 | −5.85 | −6.02 | −5.97 | −5.22 | −3.54 | −3.19 |
| 8.000 | 7.00 | −5.47 | −5.08 | −4.33 | −2.88 | −2.33 | −2.02 | −1.83 | −2.15 |
| | 7.50 | −5.60 | −5.66 | −5.64 | −4.07 | −3.24 | −2.75 | −2.25 | −2.44 |
| | 8.00 | −5.66 | −5.73 | −5.83 | −5.49 | −4.43 | −3.72 | −2.76 | −2.77 |
| | 8.30 | −5.69 | −5.76 | −5.86 | −5.98 | −5.30 | −4.46 | −3.15 | −2.99 |
| 10.000 | 7.00 | −5.12 | −4.35 | −3.60 | −2.40 | −1.97 | −1.73 | −1.68 | −2.05 |
| | 7.50 | −5.60 | −5.60 | −5.21 | −3.54 | −2.85 | −2.44 | −2.09 | −2.34 |
| | 8.00 | −5.67 | −5.74 | −5.83 | −4.88 | −3.88 | −3.27 | −2.55 | −2.65 |
| | 8.30 | −5.70 | −5.78 | −5.88 | −5.72 | −4.69 | −3.94 | −2.90 | −2.86 |

TABLE A4
Planet orbiting at 20.0 AU from a A0V star at 10.0 pc from Earth

| | | log $\mathcal{F}_\nu$ (Jy) | | | | | | | |
|---|---|---|---|---|---|---|---|---|---|
| $M/M_J$ | log $t$ | V | R | I | J | H | K | M | N |
| 0.299 | 7.00 | −6.05 | −6.13 | −6.23 | −6.41 | −6.60 | −6.80 | −5.29 | −3.94 |
| | 7.50 | −6.17 | −6.24 | −6.34 | −6.52 | −6.71 | −6.91 | −6.43 | −4.54 |
| | 8.00 | −6.25 | −6.32 | −6.42 | −6.60 | −6.79 | −6.99 | −7.37 | −5.13 |
| | 8.30 | −6.28 | −6.35 | −6.45 | −6.63 | −6.82 | −7.02 | −7.56 | −5.40 |
| 0.500 | 7.00 | −6.09 | −6.17 | −6.27 | −6.45 | −6.64 | −6.68 | −4.44 | −3.55 |
| | 7.50 | −6.18 | −6.25 | −6.35 | −6.53 | −6.73 | −6.93 | −5.58 | −4.14 |
| | 8.00 | −6.25 | −6.32 | −6.42 | −6.60 | −6.79 | −6.99 | −6.79 | −4.76 |
| | 8.30 | −6.28 | −6.35 | −6.45 | −6.63 | −6.82 | −7.02 | −7.36 | −5.12 |
| 1.000 | 7.00 | −6.14 | −6.21 | −6.31 | −6.48 | −6.11 | −5.16 | −3.43 | −3.07 |
| | 7.50 | −6.21 | −6.28 | −6.38 | −6.56 | −6.74 | −6.62 | −4.34 | −3.56 |
| | 8.00 | −6.26 | −6.33 | −6.43 | −6.61 | −6.80 | −7.00 | −5.53 | −4.16 |
| | 8.30 | −6.28 | −6.36 | −6.46 | −6.64 | −6.83 | −7.03 | −6.37 | −4.57 |
| 2.000 | 7.00 | −6.15 | −6.23 | −6.33 | −5.89 | −4.67 | −3.89 | −2.79 | −2.74 |
| | 7.50 | −6.21 | −6.29 | −6.39 | −6.56 | −6.19 | −5.24 | −3.51 | −3.14 |
| | 8.00 | −6.26 | −6.33 | −6.43 | −6.61 | −6.80 | −6.75 | −4.47 | −3.65 |
| | 8.30 | −6.28 | −6.35 | −6.45 | −6.63 | −6.82 | −7.01 | −5.19 | −4.01 |
| 3.000 | 7.00 | −6.15 | −6.22 | −6.31 | −4.98 | −3.91 | −3.27 | −2.49 | −2.56 |
| | 7.50 | −6.21 | −6.28 | −6.38 | −6.39 | −5.32 | −4.44 | −3.10 | −2.93 |
| | 8.00 | −6.25 | −6.32 | −6.42 | −6.60 | −6.68 | −5.99 | −3.93 | −3.38 |
| | 8.30 | −6.27 | −6.35 | −6.45 | −6.63 | −6.81 | −6.81 | −4.55 | −3.70 |
| 4.000 | 7.00 | −6.14 | −6.21 | −6.15 | −4.32 | −3.41 | −2.87 | −2.28 | −2.44 |
| | 7.50 | −6.20 | −6.28 | −6.37 | −5.89 | −4.68 | −3.90 | −2.83 | −2.78 |
| | 8.00 | −6.25 | −6.32 | −6.42 | −6.59 | −6.26 | −5.31 | −3.56 | −3.19 |
| | 8.30 | −6.27 | −6.34 | −6.44 | −6.62 | −6.77 | −6.30 | −4.12 | −3.49 |
| 6.000 | 7.00 | −6.12 | −5.94 | −5.22 | −3.46 | −2.76 | −2.36 | −2.01 | −2.27 |
| | 7.50 | −6.20 | −6.27 | −6.34 | −4.85 | −3.83 | −3.21 | −2.48 | −2.58 |
| | 8.00 | −6.25 | −6.32 | −6.42 | −6.38 | −5.26 | −4.40 | −3.10 | −2.95 |
| | 8.30 | −6.28 | −6.35 | −6.45 | −6.62 | −6.21 | −5.25 | −3.54 | −3.19 |
| 8.000 | 7.00 | −5.94 | −5.20 | −4.35 | −2.88 | −2.33 | −2.02 | −1.83 | −2.15 |
| | 7.50 | −6.20 | −6.24 | −5.99 | −4.08 | −3.25 | −2.75 | −2.25 | −2.44 |
| | 8.00 | −6.26 | −6.33 | −6.43 | −5.61 | −4.44 | −3.72 | −2.76 | −2.77 |
| | 8.30 | −6.29 | −6.36 | −6.46 | −6.44 | −5.33 | −4.46 | −3.15 | −2.99 |
| 10.000 | 7.00 | −5.28 | −4.37 | −3.60 | −2.40 | −1.97 | −1.73 | −1.68 | −2.05 |
| | 7.50 | −6.19 | −6.02 | −5.31 | −3.54 | −2.85 | −2.44 | −2.09 | −2.34 |
| | 8.00 | −6.27 | −6.34 | −6.41 | −4.90 | −3.88 | −3.28 | −2.55 | −2.65 |
| | 8.30 | −6.30 | −6.38 | −6.48 | −5.91 | −4.70 | −3.94 | −2.90 | −2.86 |

TABLE A5
Planet orbiting at 2.6 AU from a M5V star at 10.0 pc from Earth

| $M/M_J$ | $\log t$ | V | R | I | J | H | K | M | N |
|---|---|---|---|---|---|---|---|---|---|
| | | | | $\log \mathcal{F}_\nu$ (Jy) | | | | | |
| 0.299 | 7.50 | −9.31 | −8.87 | −8.31 | −7.72 | −7.70 | −7.80 | −6.70 | −4.65 |
| | 8.00 | −9.75 | −9.30 | −8.75 | −8.15 | −8.13 | −8.23 | −8.26 | −5.48 |
| | 8.50 | −10.04 | −9.59 | −9.04 | −8.45 | −8.42 | −8.52 | −9.02 | −6.47 |
| | 9.00 | −10.10 | −9.66 | −9.10 | −8.51 | −8.49 | −8.59 | −9.10 | −7.85 |
| | 9.50 | −10.16 | −9.71 | −9.16 | −8.57 | −8.54 | −8.64 | −9.15 | −9.37 |
| 0.500 | 7.50 | −9.31 | −8.87 | −8.32 | −7.72 | −7.70 | −7.80 | −5.67 | −4.17 |
| | 8.00 | −9.72 | −9.28 | −8.73 | −8.13 | −8.11 | −8.21 | −7.18 | −4.91 |
| | 8.50 | −10.00 | −9.55 | −9.00 | −8.40 | −8.38 | −8.48 | −8.72 | −5.78 |
| | 9.00 | −10.06 | −9.61 | −9.06 | −8.47 | −8.44 | −8.54 | −9.05 | −6.95 |
| | 9.50 | −10.10 | −9.66 | −9.11 | −8.51 | −8.49 | −8.59 | −9.10 | −8.49 |
| | 9.70 | −10.12 | −9.68 | −9.12 | −8.53 | −8.50 | −8.61 | −9.12 | −9.15 |
| 1.000 | 7.50 | −9.34 | −8.89 | −8.34 | −7.75 | −7.71 | −7.12 | −4.37 | −3.56 |
| | 8.00 | −9.73 | −9.28 | −8.73 | −8.14 | −8.11 | −8.20 | −5.63 | −4.19 |
| | 8.50 | −9.98 | −9.53 | −8.98 | −8.39 | −8.36 | −8.47 | −7.26 | −4.97 |
| | 9.00 | −10.02 | −9.57 | −9.02 | −8.43 | −8.40 | −8.51 | −8.83 | −5.90 |
| | 9.50 | −10.06 | −9.61 | −9.06 | −8.47 | −8.44 | −8.54 | −9.05 | −7.15 |
| | 9.70 | −10.07 | −9.63 | −9.07 | −8.48 | −8.46 | −8.56 | −9.07 | −7.74 |
| 2.000 | 7.50 | −9.34 | −8.90 | −8.34 | −7.71 | −6.80 | −5.41 | −3.51 | −3.14 |
| | 8.00 | −9.72 | −9.28 | −8.72 | −8.13 | −8.09 | −7.37 | −4.50 | −3.65 |
| | 8.50 | −9.97 | −9.52 | −8.97 | −8.38 | −8.35 | −8.45 | −5.85 | −4.31 |
| | 9.00 | −10.00 | −9.55 | −9.00 | −8.41 | −8.38 | −8.49 | −7.39 | −5.04 |
| | 9.50 | −10.03 | −9.58 | −9.03 | −8.44 | −8.41 | −8.52 | −8.90 | −5.99 |
| | 9.70 | −10.04 | −9.60 | −9.04 | −8.45 | −8.43 | −8.53 | −9.02 | −6.44 |
| 3.000 | 7.50 | −9.34 | −8.89 | −8.34 | −7.15 | −5.69 | −4.54 | −3.10 | −2.92 |
| | 8.00 | −9.72 | −9.27 | −8.72 | −8.12 | −7.73 | −6.26 | −3.94 | −3.37 |
| | 8.50 | −9.96 | −9.52 | −8.96 | −8.37 | −8.34 | −8.23 | −5.08 | −3.95 |
| | 9.00 | −9.99 | −9.54 | −8.99 | −8.40 | −8.37 | −8.48 | −6.48 | −4.62 |
| | 9.50 | −10.02 | −9.57 | −9.02 | −8.43 | −8.40 | −8.51 | −8.20 | −5.45 |
| | 9.70 | −10.03 | −9.58 | −9.03 | −8.44 | −8.41 | −8.52 | −8.75 | −5.81 |
| 4.000 | 7.50 | −9.33 | −8.88 | −8.30 | −6.28 | −4.94 | −3.97 | −2.83 | −2.77 |
| | 8.00 | −9.72 | −9.27 | −8.72 | −8.06 | −6.91 | −5.49 | −3.57 | −3.19 |
| | 8.50 | −9.96 | −9.51 | −8.96 | −8.37 | −8.32 | −7.53 | −4.58 | −3.71 |
| | 9.00 | −9.99 | −9.54 | −8.99 | −8.40 | −8.37 | −8.47 | −5.90 | −4.35 |
| | 9.50 | −10.02 | −9.57 | −9.02 | −8.42 | −8.40 | −8.50 | −7.47 | −5.09 |
| | 9.70 | −10.03 | −9.58 | −9.03 | −8.43 | −8.41 | −8.51 | −8.15 | −5.42 |
| 6.000 | 7.50 | −9.31 | −8.51 | −7.44 | −5.03 | −3.98 | −3.25 | −2.48 | −2.58 |
| | 8.00 | −9.72 | −9.27 | −8.71 | −7.12 | −5.62 | −4.50 | −3.10 | −2.94 |
| | 8.50 | −9.97 | −9.52 | −8.97 | −8.36 | −7.69 | −6.14 | −3.90 | −3.38 |
| | 9.00 | −10.00 | −9.55 | −9.00 | −8.41 | −8.38 | −8.23 | −5.08 | −3.97 |
| | 9.50 | −10.02 | −9.58 | −9.02 | −8.43 | −8.41 | −8.51 | −6.48 | −4.64 |
| | 9.70 | −10.03 | −9.59 | −9.03 | −8.44 | −8.42 | −8.52 | −7.11 | −4.93 |
| 8.000 | 7.50 | −8.82 | −7.33 | −6.22 | −4.19 | −3.34 | −2.77 | −2.25 | −2.44 |
| | 8.00 | −9.73 | −9.23 | −8.45 | −5.90 | −4.66 | −3.78 | −2.76 | −2.76 |
| | 8.50 | −9.98 | −9.54 | −8.98 | −8.07 | −6.53 | −5.20 | −3.46 | −3.16 |
| | 9.00 | −10.02 | −9.57 | −9.02 | −8.43 | −8.37 | −7.41 | −4.55 | −3.72 |
| | 9.50 | −10.04 | −9.60 | −9.04 | −8.45 | −8.42 | −8.52 | −6.01 | −4.43 |
| | 9.70 | −10.05 | −9.60 | −9.05 | −8.46 | −8.43 | −8.54 | −6.49 | −4.65 |
| 10.000 | 7.50 | −7.76 | −6.36 | −5.35 | −3.61 | −2.90 | −2.44 | −2.09 | −2.34 |
| | 8.00 | −9.70 | −8.66 | −7.50 | −5.07 | −4.03 | −3.31 | −2.55 | −2.64 |
| | 8.50 | −10.00 | −9.55 | −8.99 | −7.20 | −5.68 | −4.56 | −3.17 | −3.01 |
| | 9.00 | −10.04 | −9.59 | −9.04 | −8.44 | −8.12 | −6.63 | −4.17 | −3.54 |
| | 9.50 | −10.06 | −9.61 | −9.06 | −8.47 | −8.44 | −8.52 | −5.65 | −4.27 |
| | 9.70 | −10.06 | −9.62 | −9.07 | −8.47 | −8.45 | −8.55 | −6.07 | −4.47 |

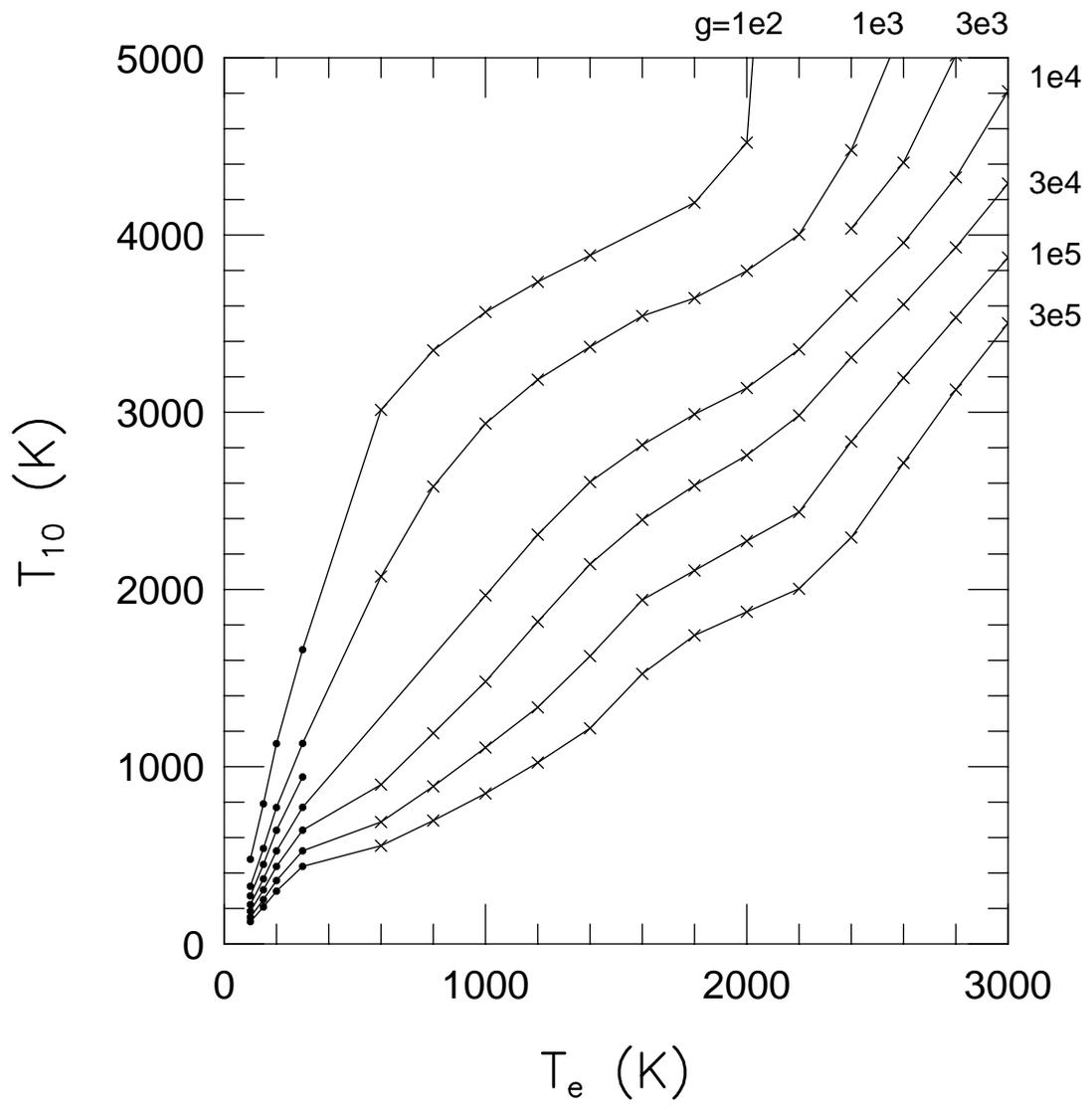

Fig. 1

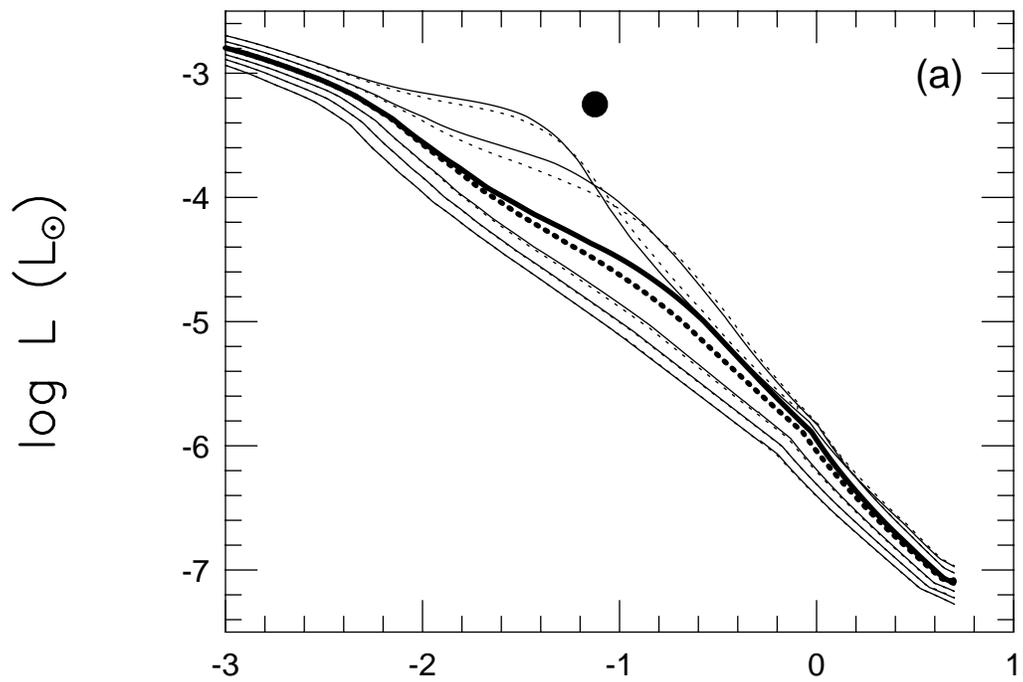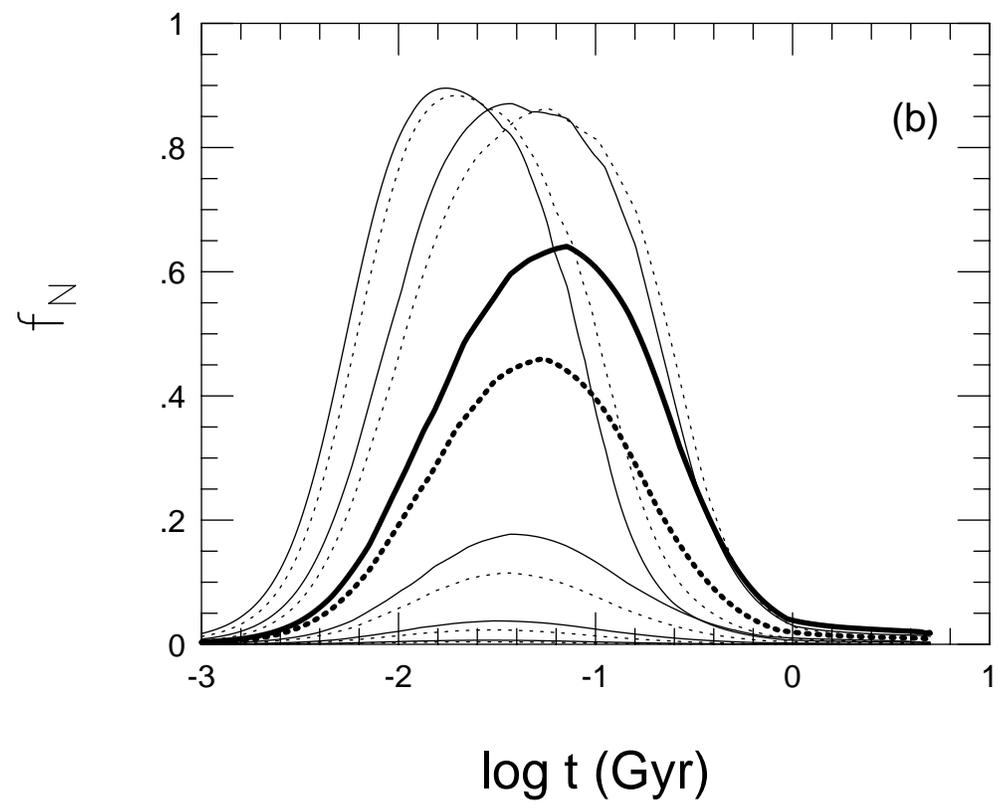

Fig. 2 (a), (b)

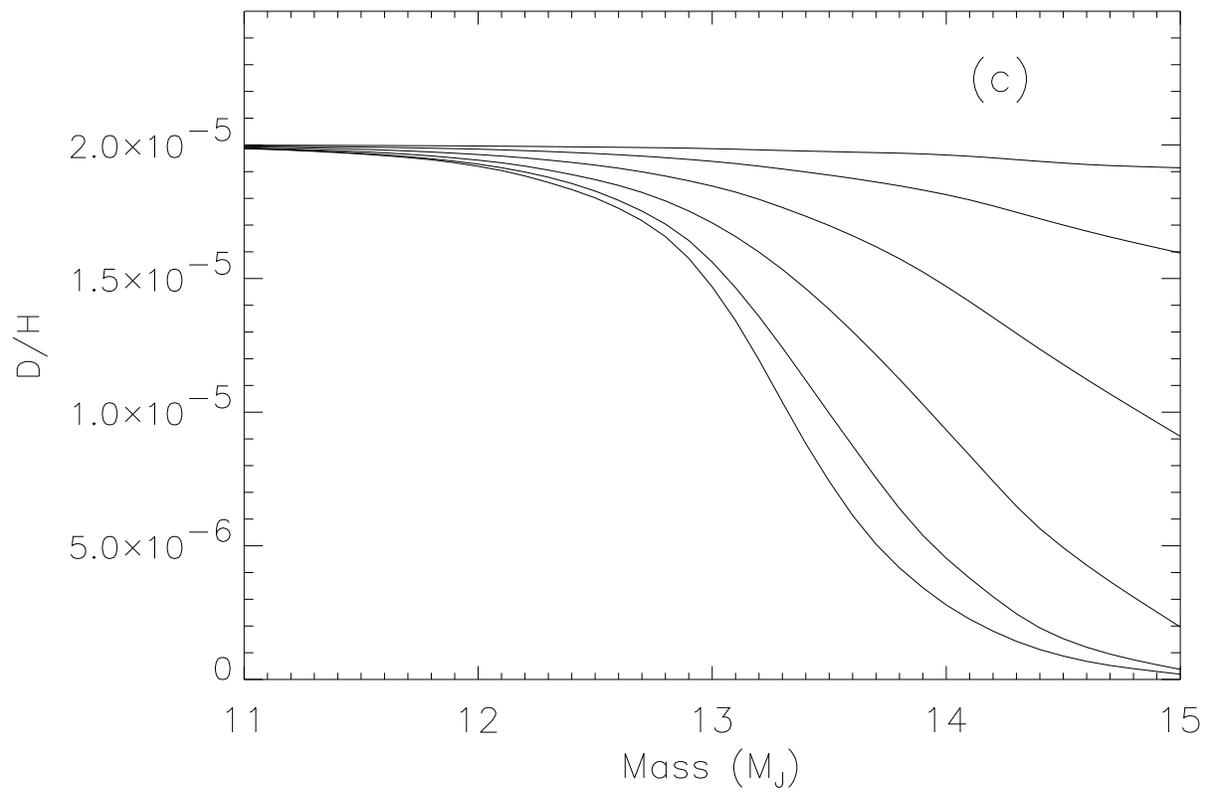

Fig. 2(c)

(a)

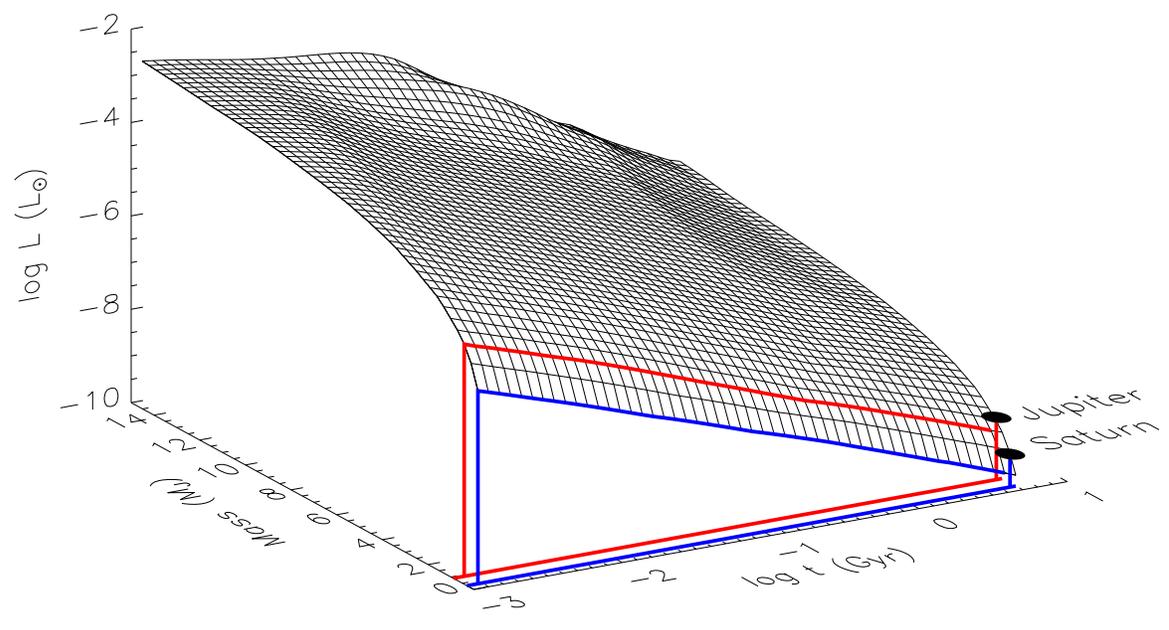

Fig. 3 (a)

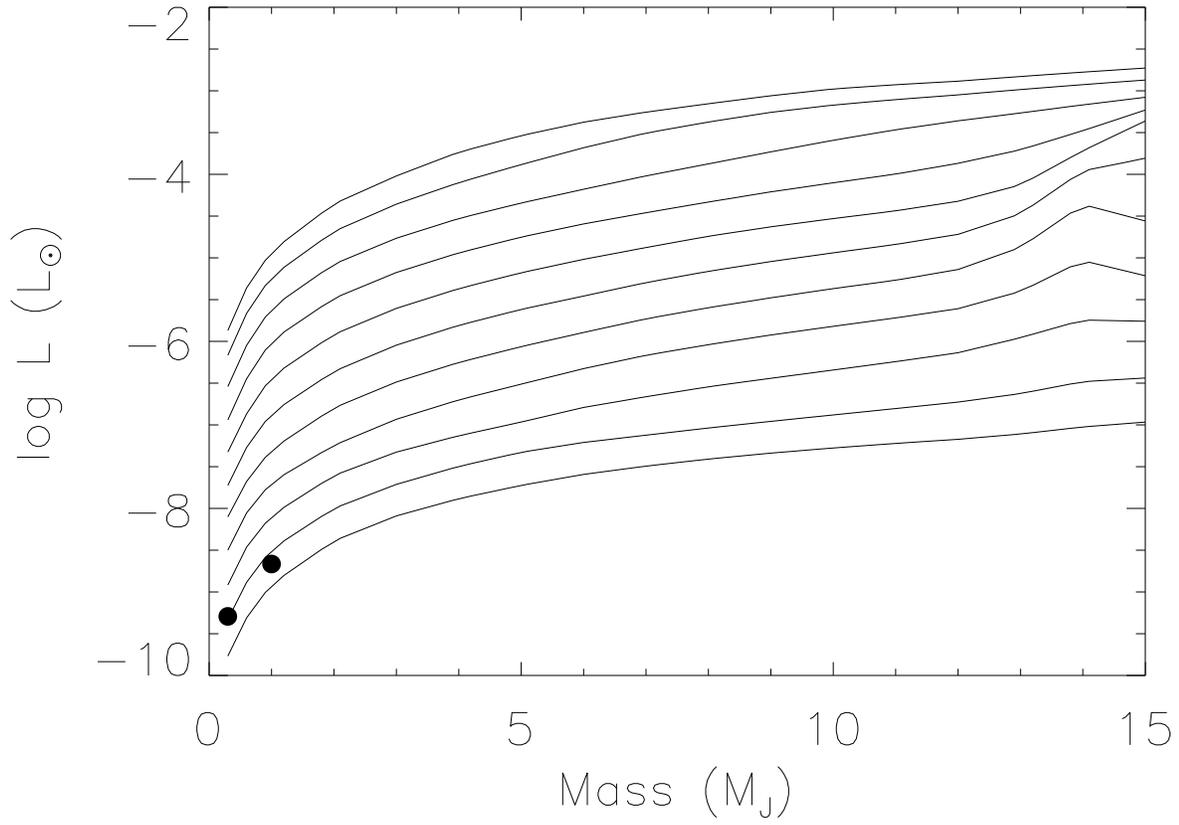

Fig. 3 (b)

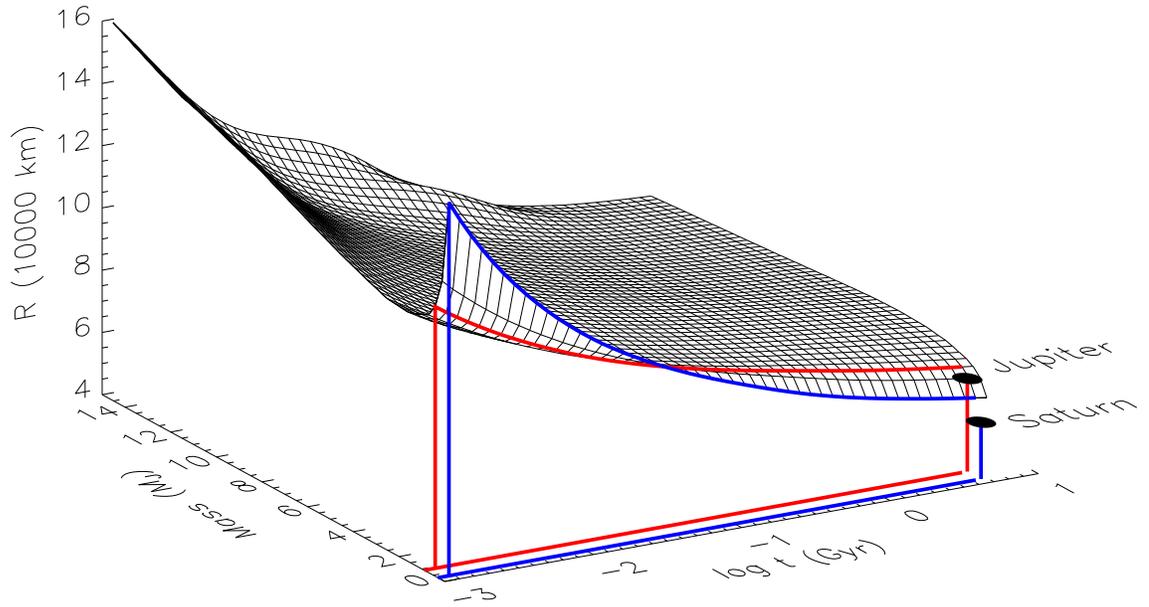

Fig. 4 (a)

(b)

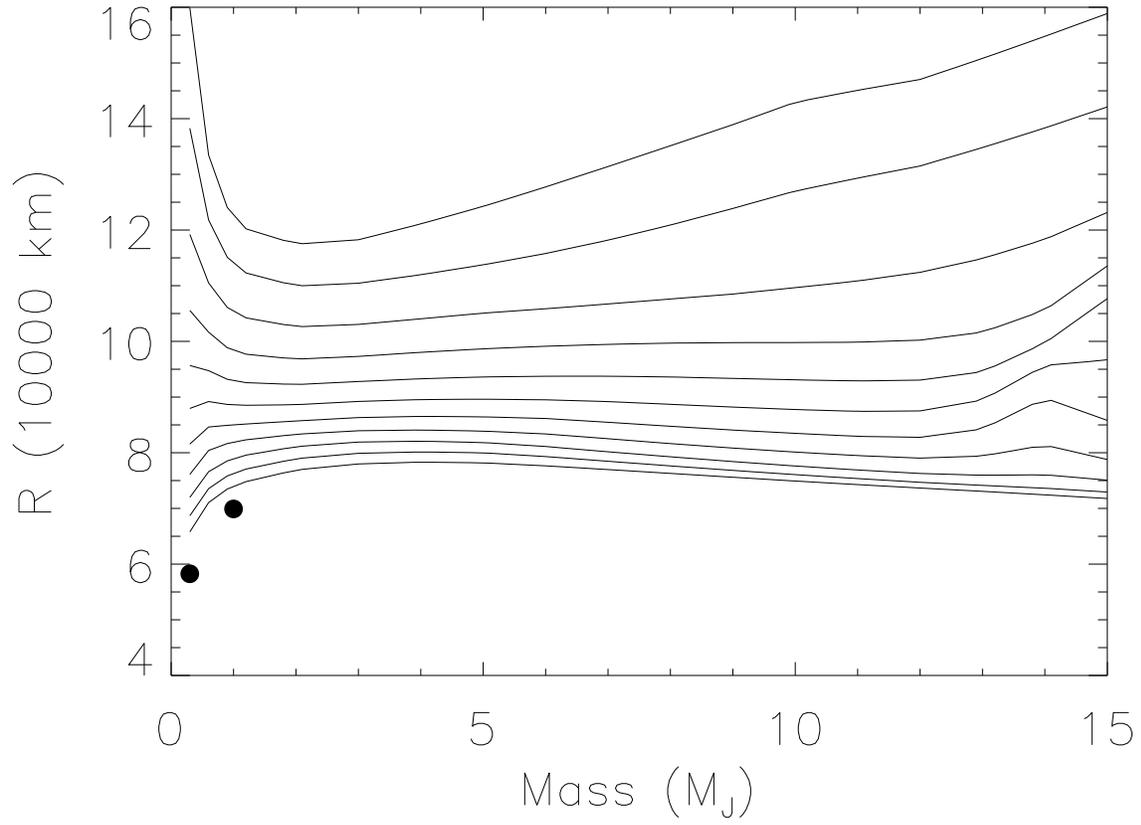

Fig. 4 (b)

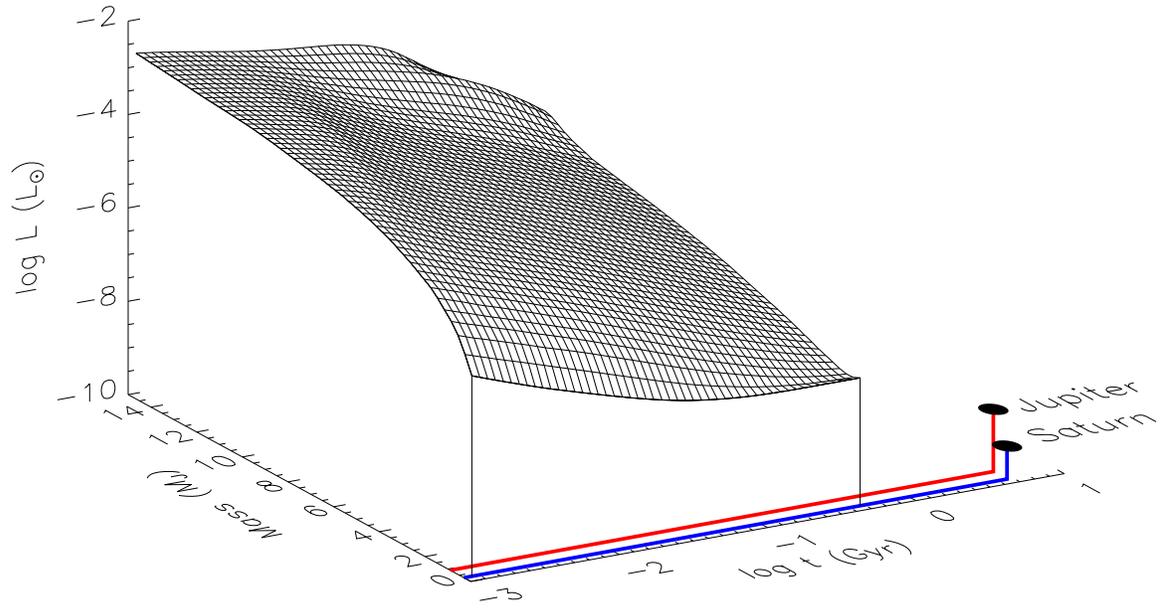

Fig. 5 (a)

(b)

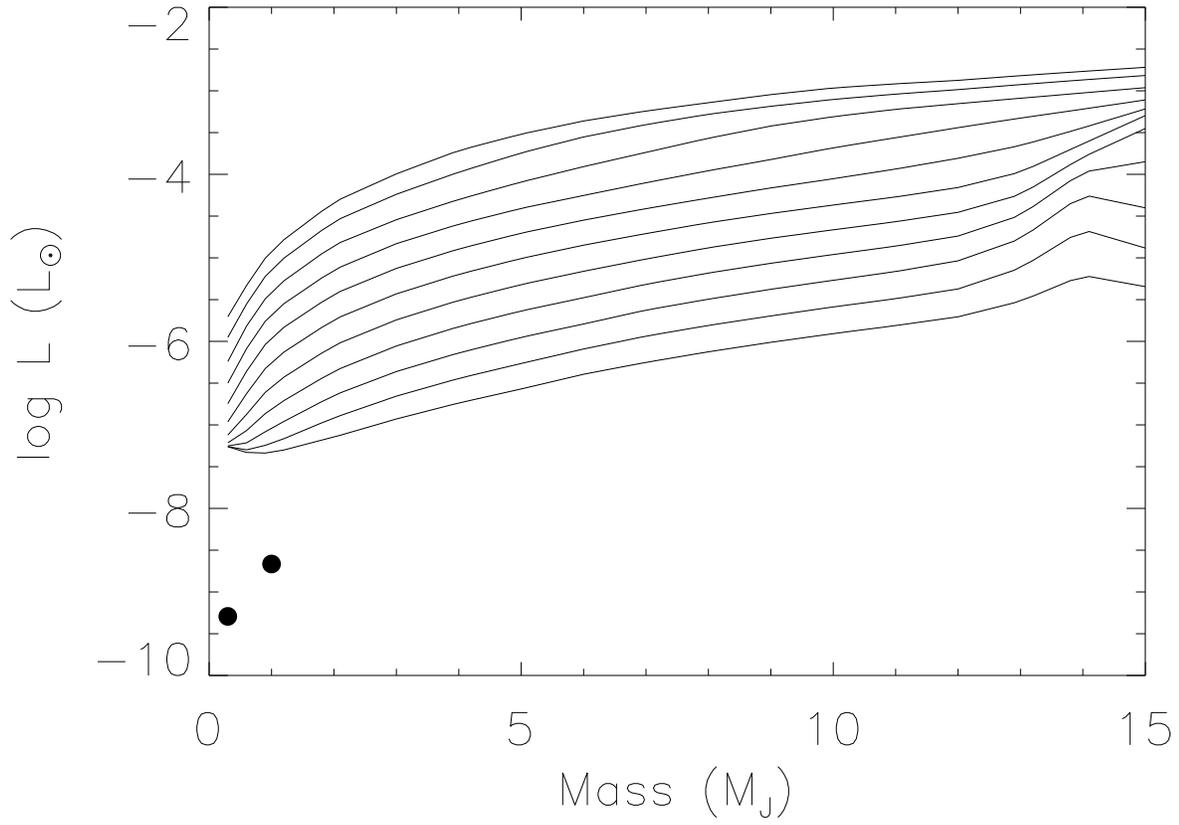

Fig. 5 (b)

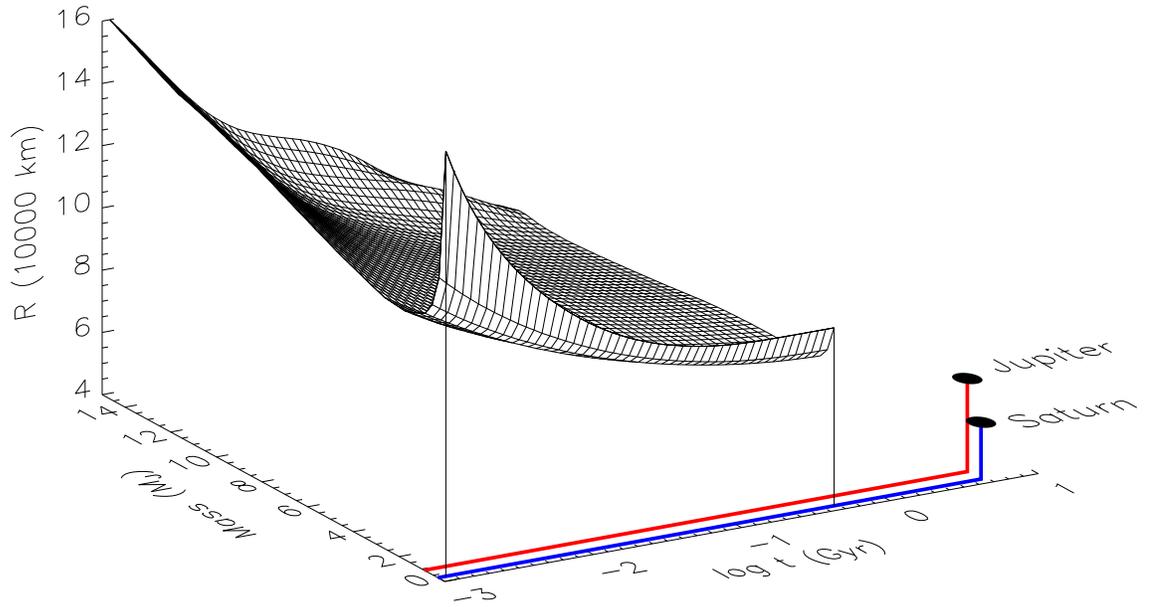

Fig. 6(a)

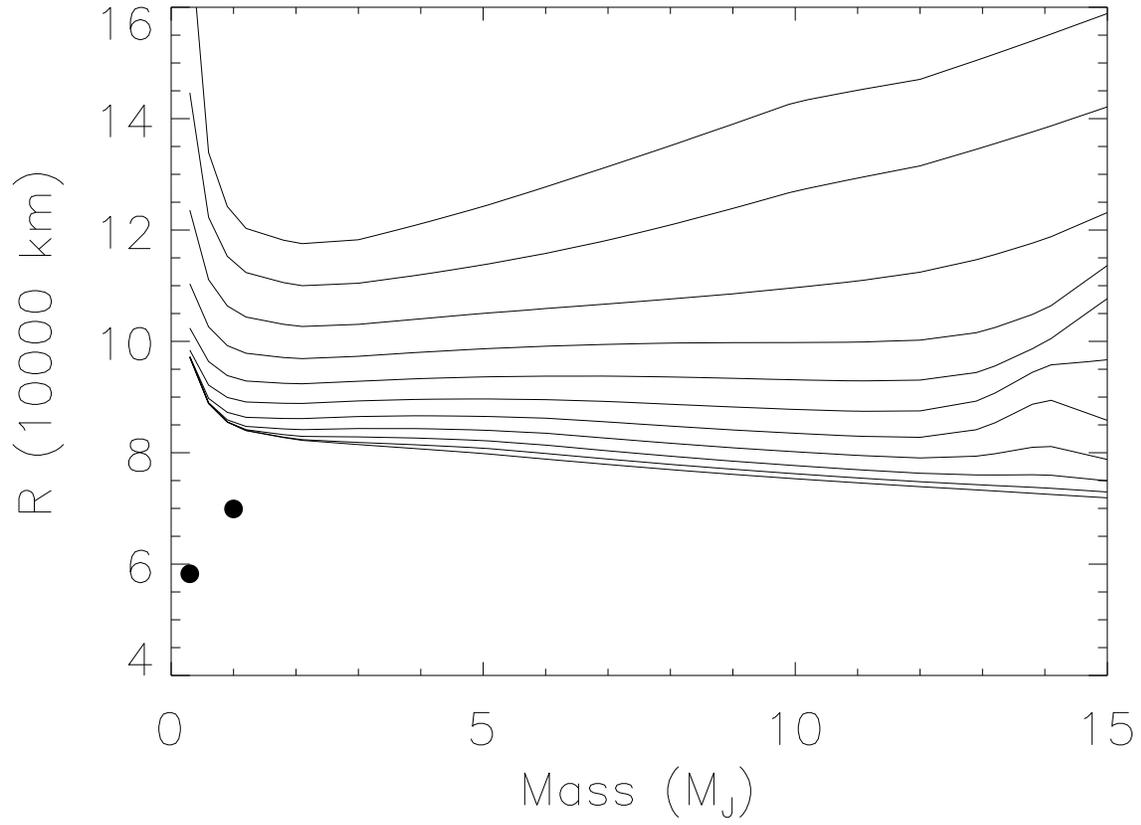

Fig. 6 (b)

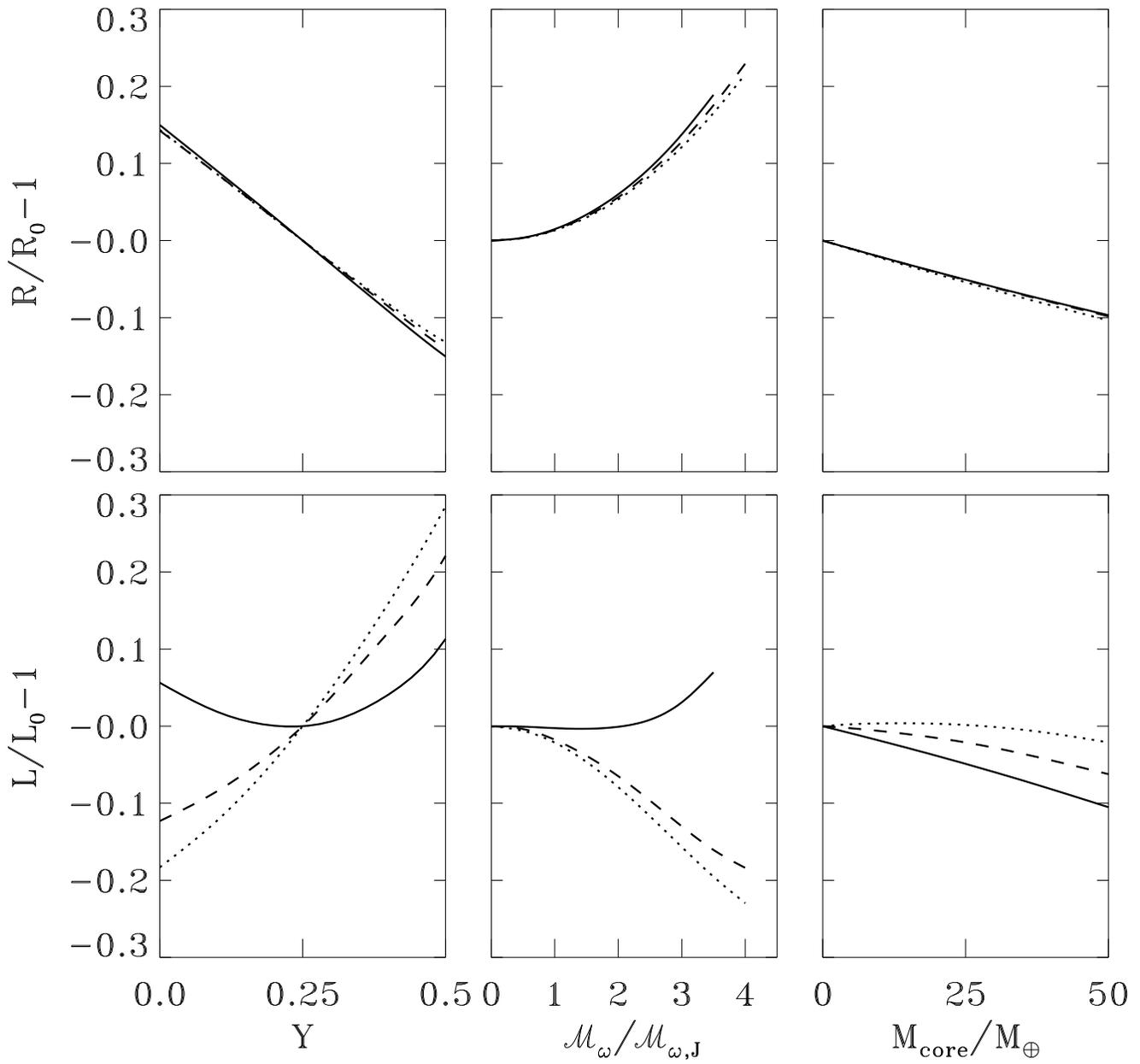

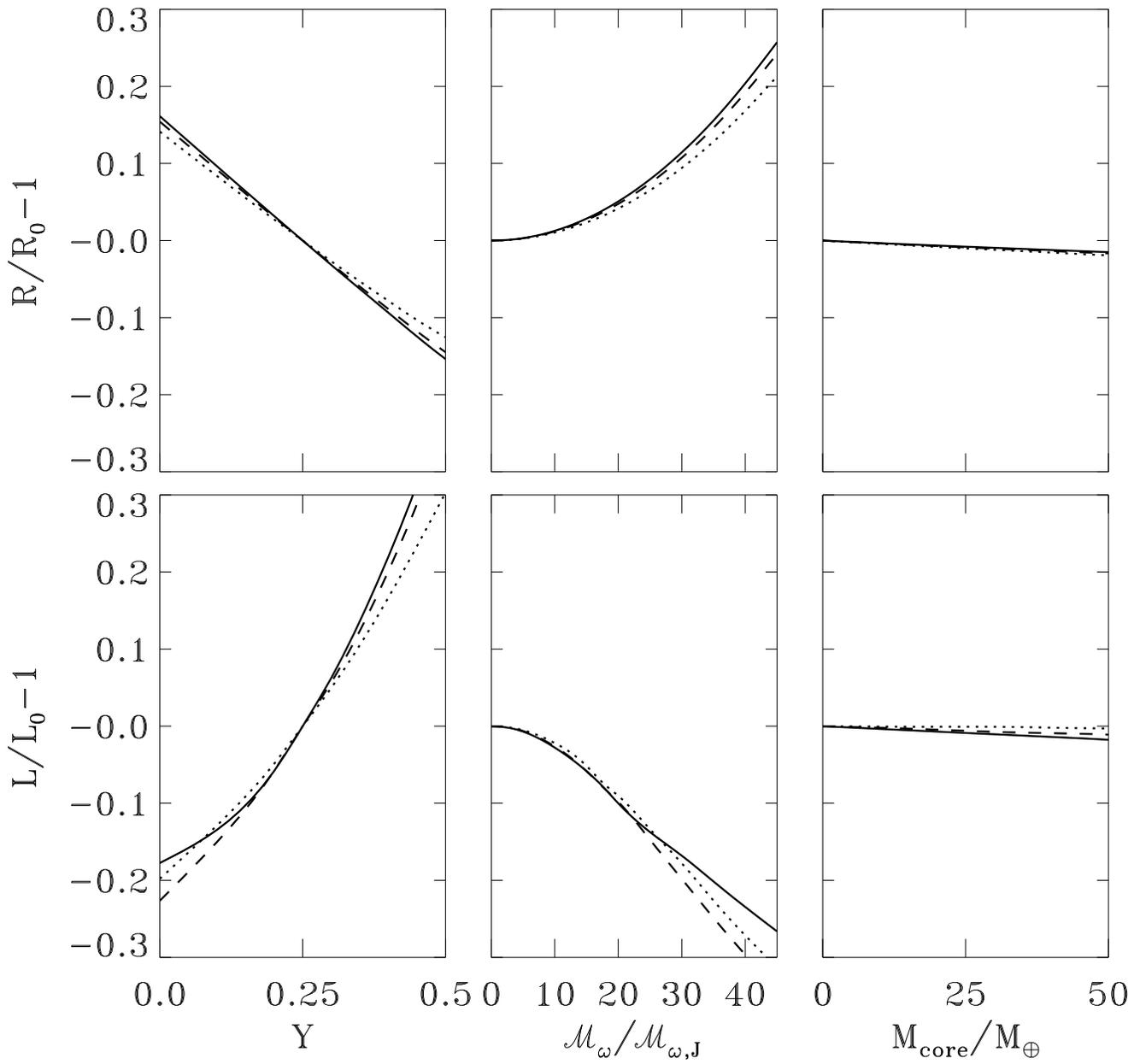

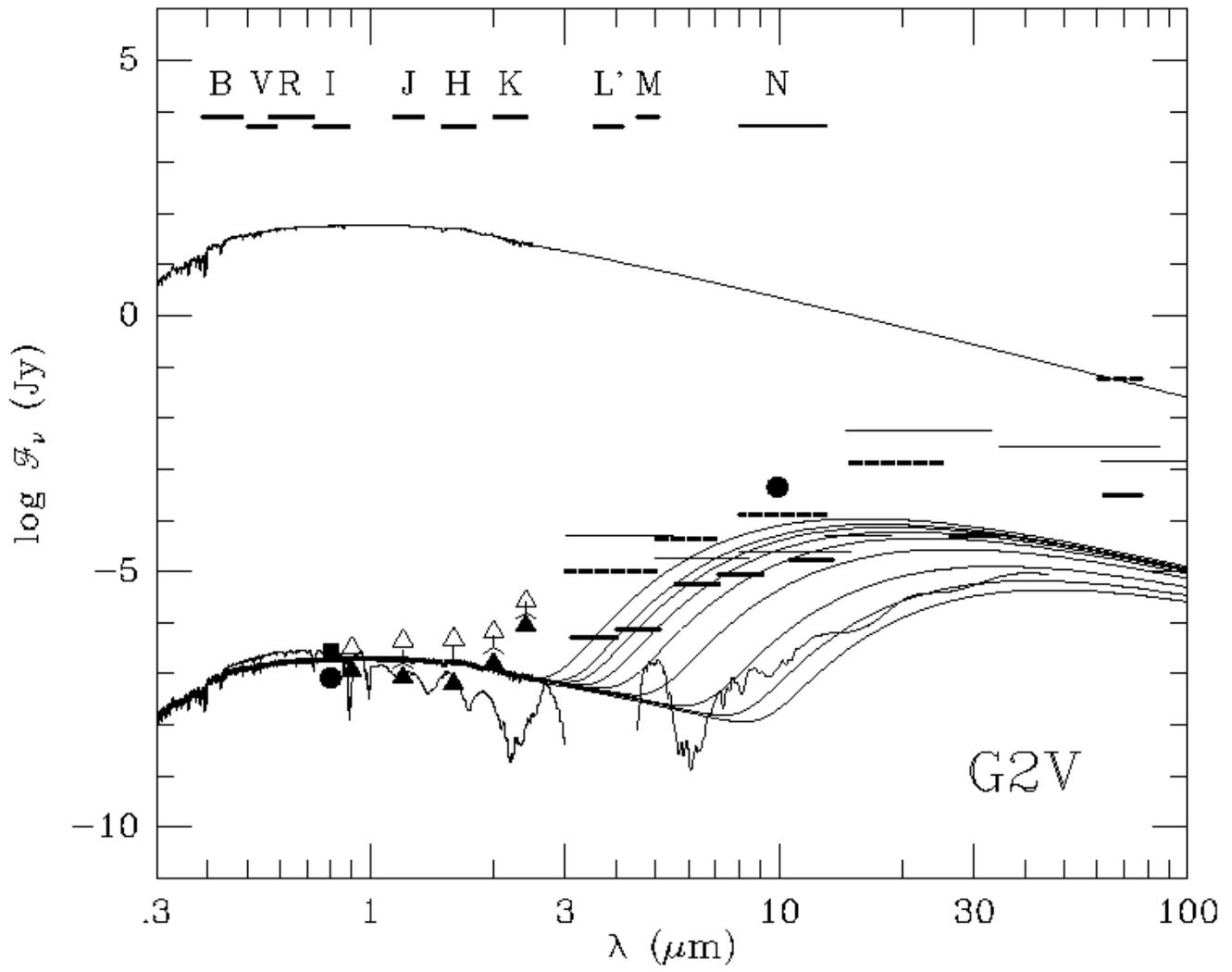

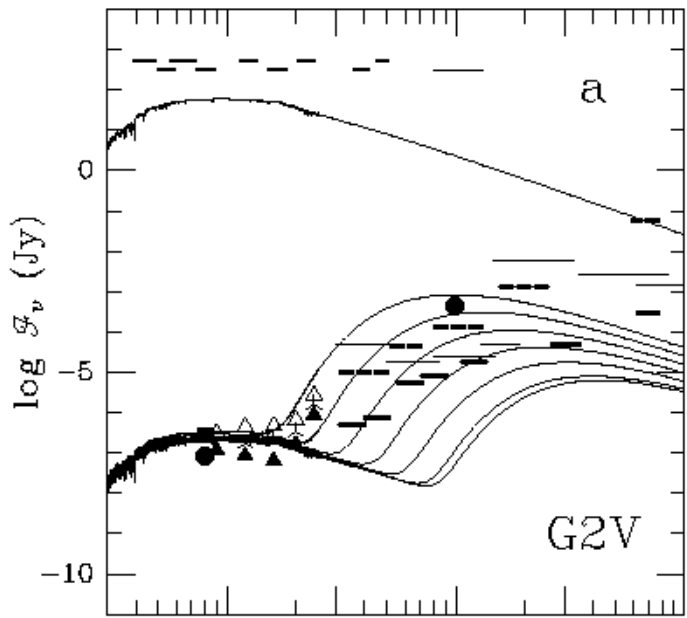
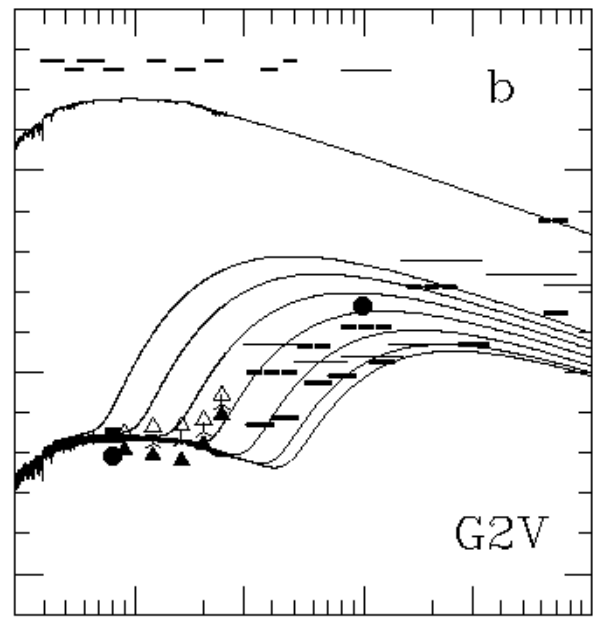
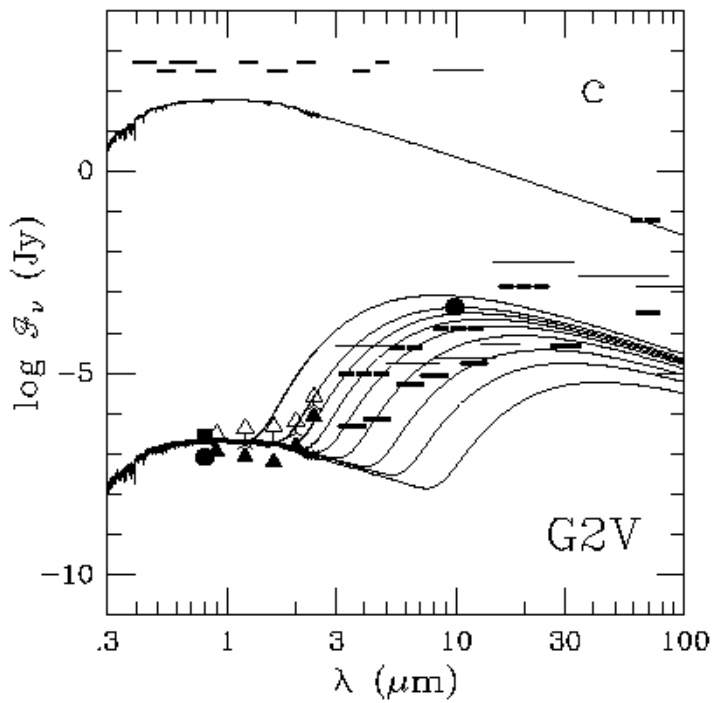
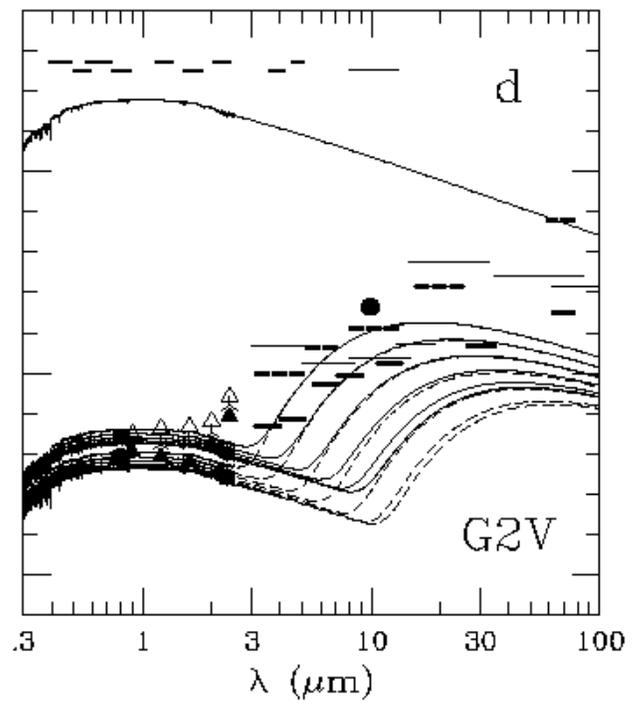

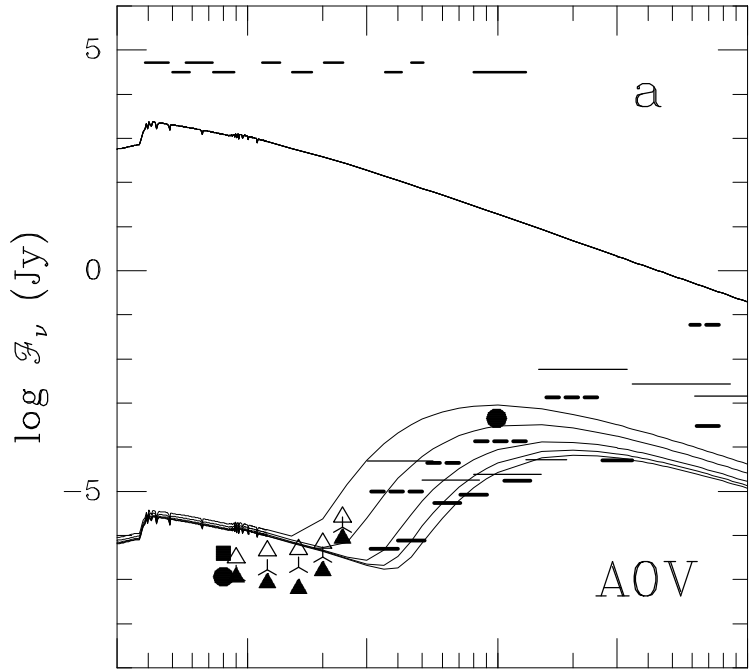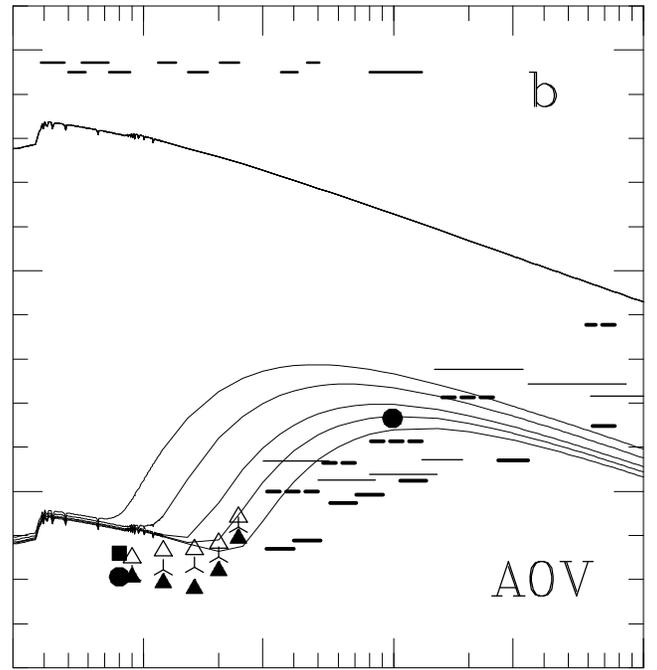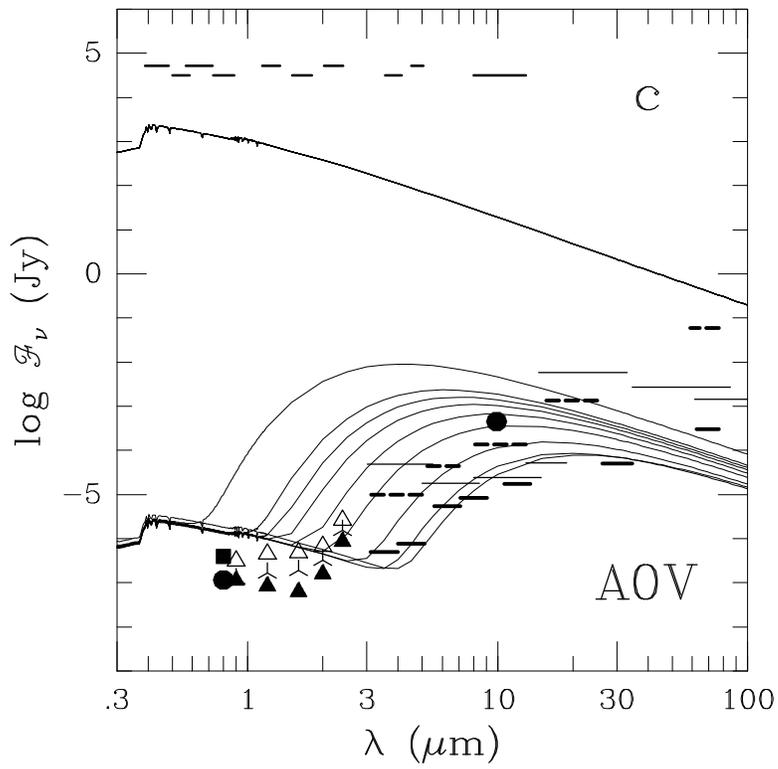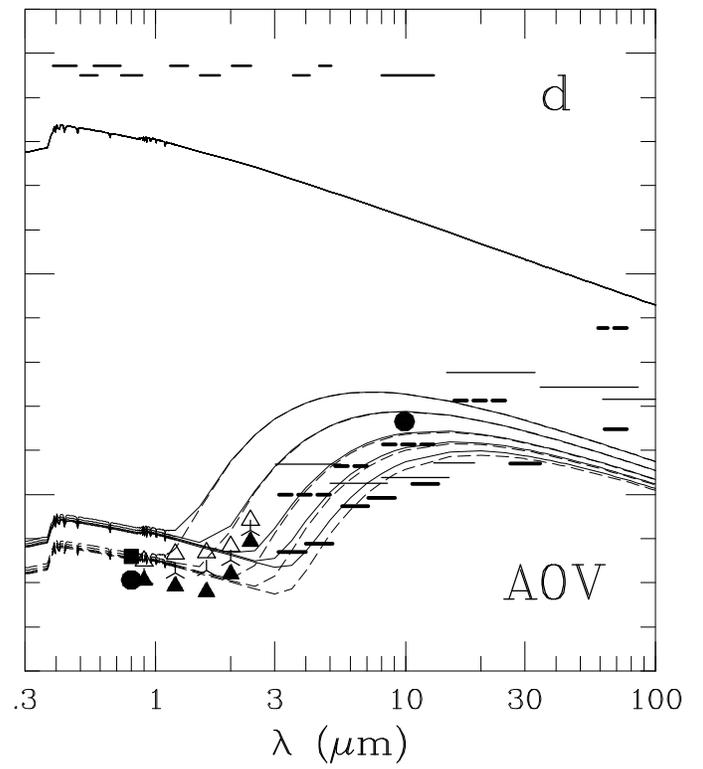

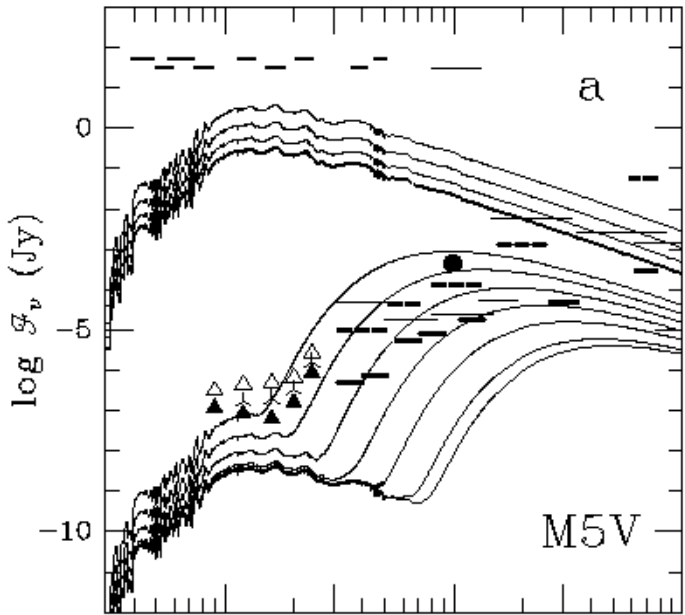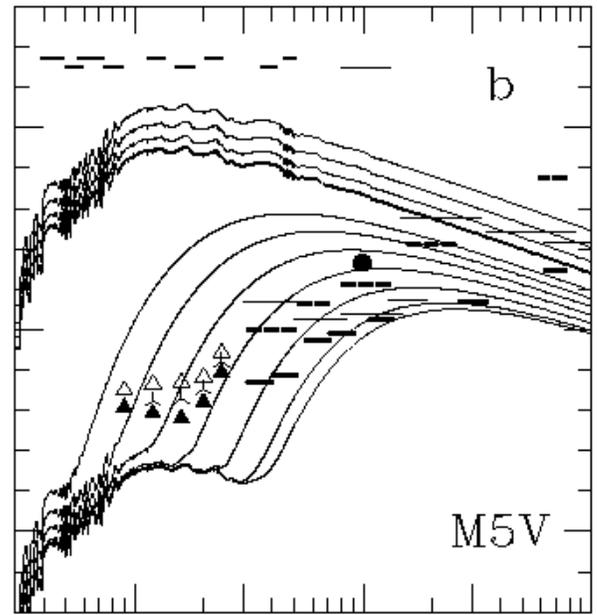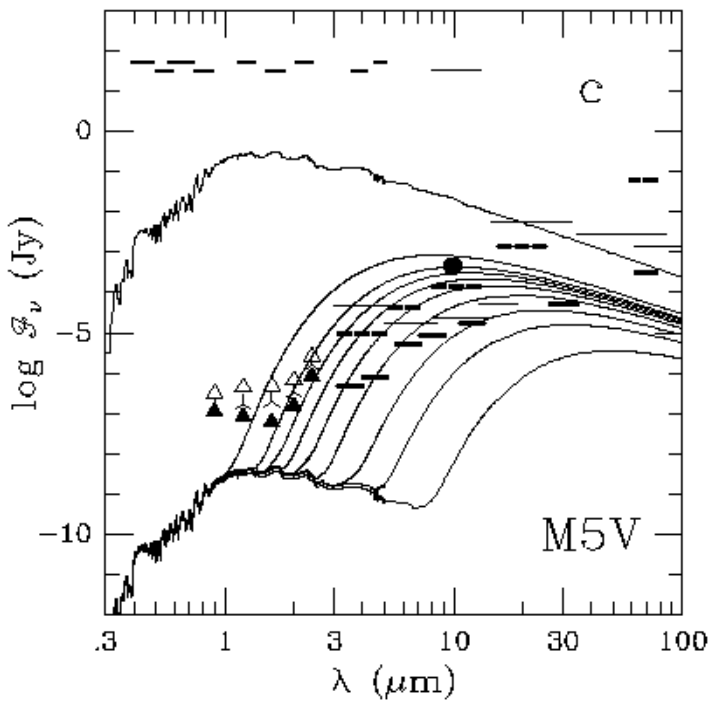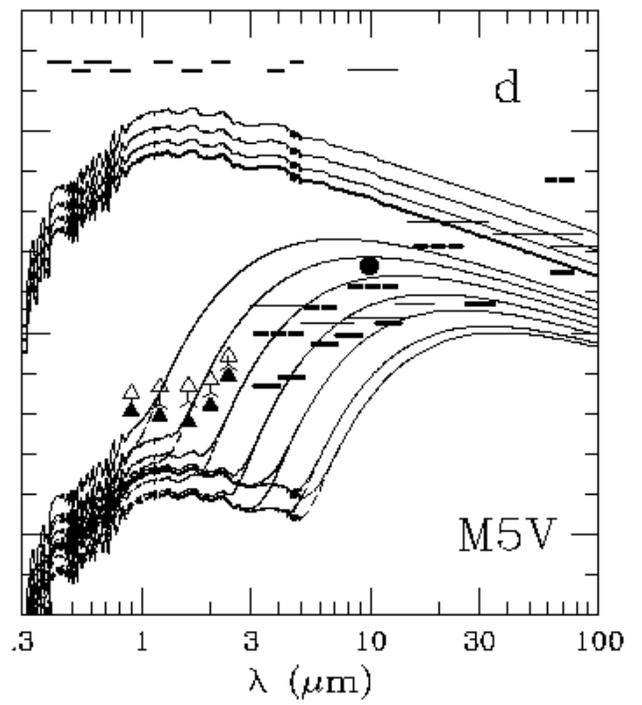

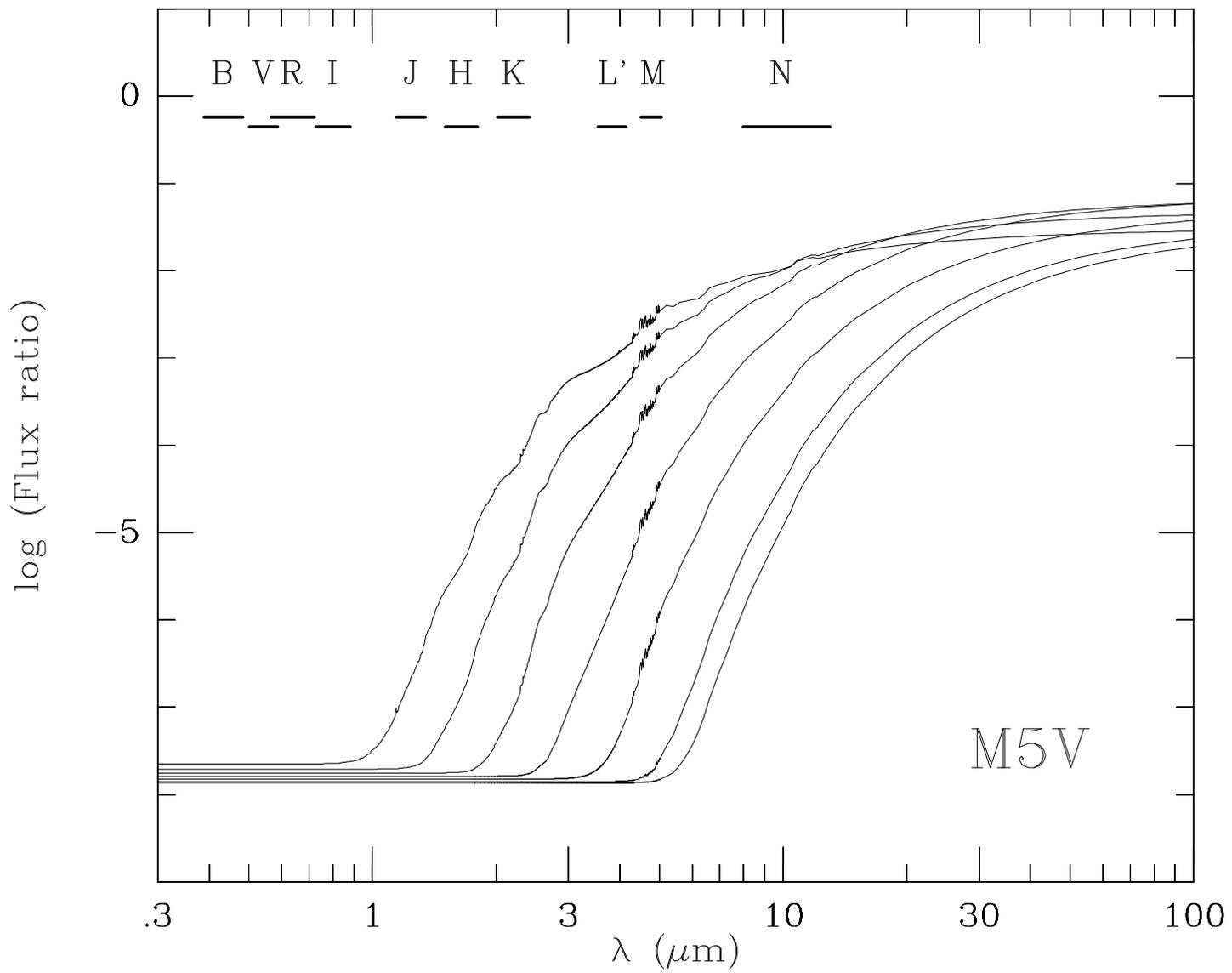

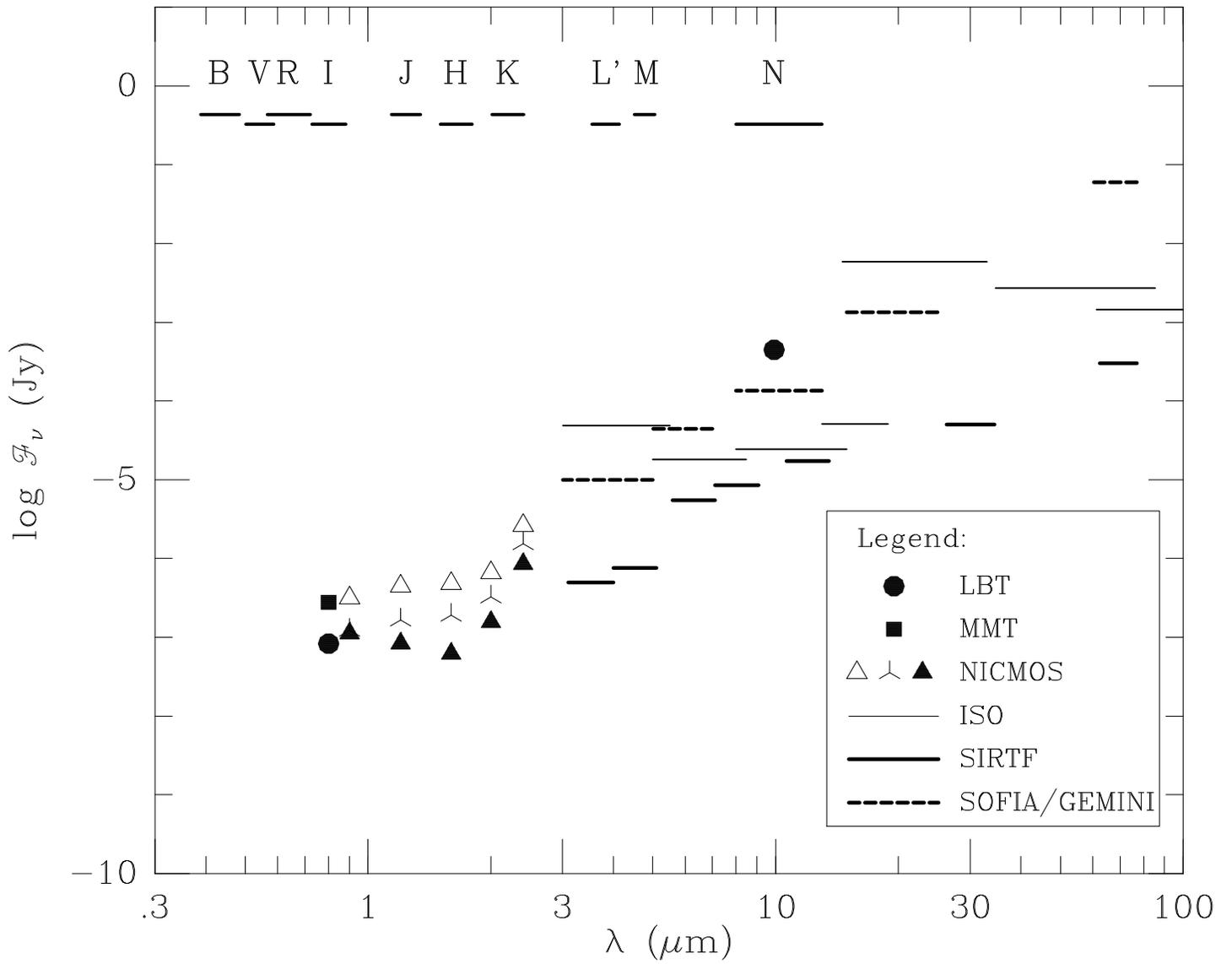

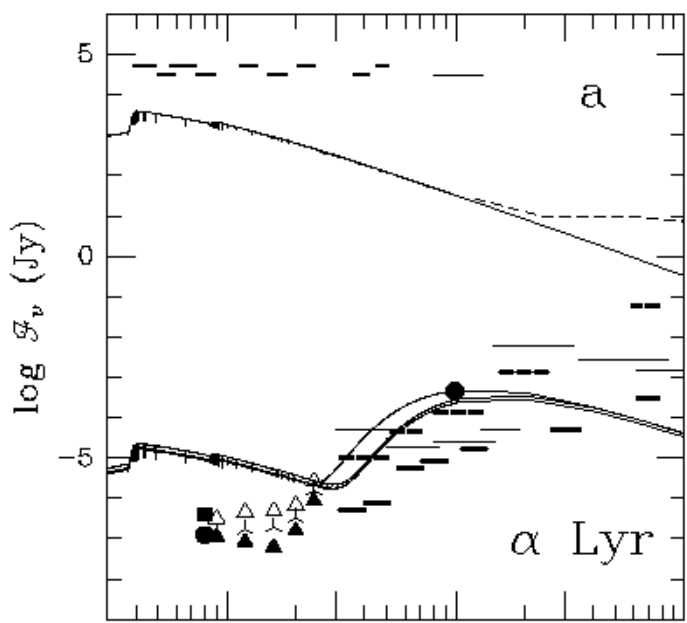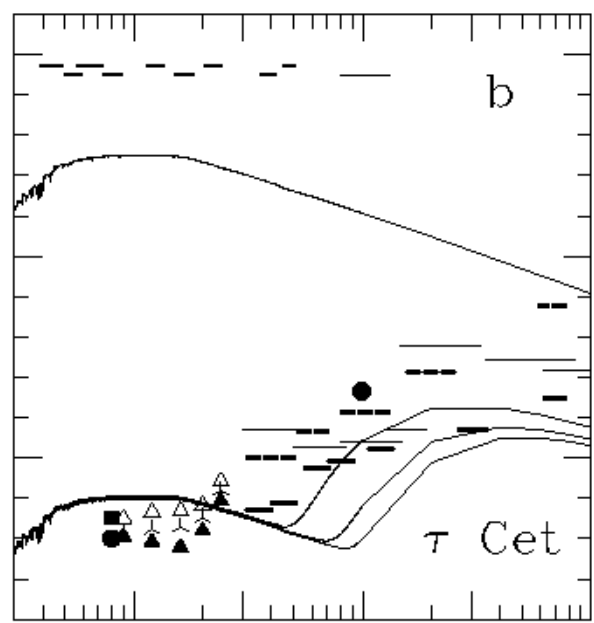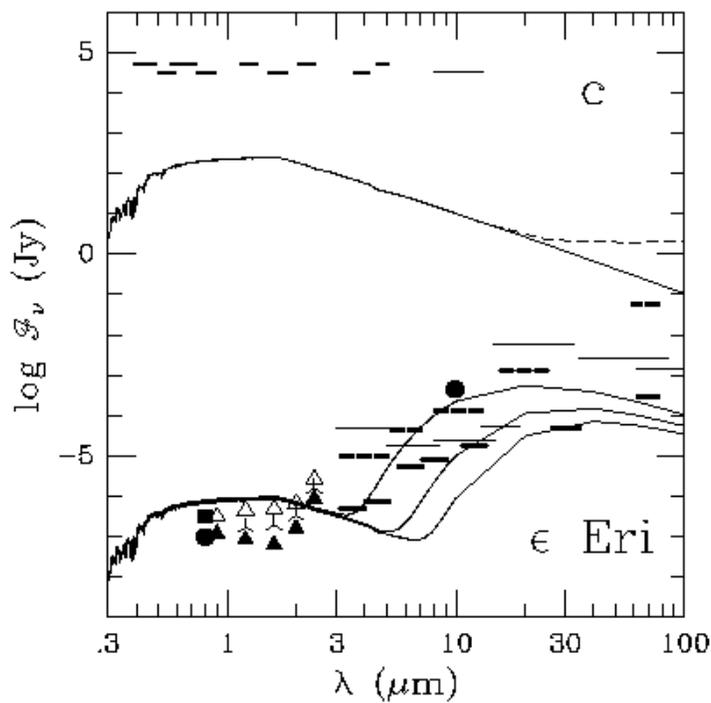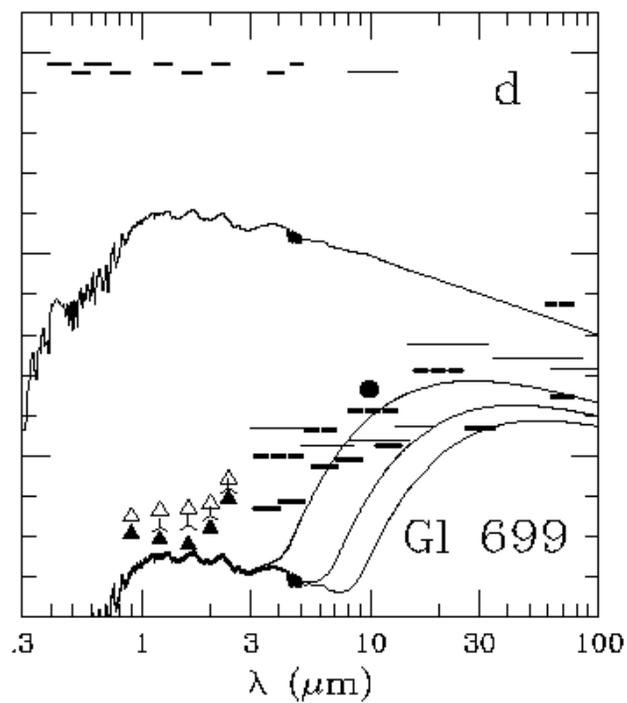